\documentclass[12pt]{article}

\usepackage{color}
\usepackage[usenames,dvipsnames]{xcolor}
\usepackage[colorlinks=true,linkcolor=black,citecolor=Maroon]{hyperref}%Maroon
\usepackage{graphicx}
\usepackage{subfigure}
\usepackage{amsmath}
\usepackage{amsfonts}
\usepackage{amssymb}
\usepackage[T1]{fontenc}

\newcommand{\comment}[1]{}

\def\bea{\begin{eqnarray}}
\def\eea{\end{eqnarray}}
\def\be{\begin{equation}}
\def\ee{\end{equation}}

\def\g{\gamma}
\def\l{\lambda}

\def\d{{\partial}}

\def\vep{\varepsilon}
\def\vphi{\varphi}
\def\T{\mathcal{T}}
\def\mbf{\mathbf}

\def\O{\mathcal{O}}

\begin{document}

\begin{center}

{\bf \Large Black hole discharge in massive electrodynamics\\[12 pt] and black hole disappearance in massive gravity}

\vskip 1 cm
{\bf Mehrdad Mirbabayi, Andrei Gruzinov}

\vskip 0.5 cm
{CCPP, Physics department, NYU, 4 Washington place, New York, NY 10003}

\end{center}

\begin{abstract}

We define and calculate the {\it discharge mode} for a Schwarzschild black hole in massive electrodynamics. For small photon mass, the discharge mode describes the decay of the electric field of a charged star collapsing into a black hole. We argue that a similar ``discharge of mass'' occurs in massive gravity and leads to a strange process of black hole disappearance.

~~

\end{abstract}

\parskip 0.2 cm

\section{Introduction and Estimates}

In massive electrodynamics, a charged star collapsing into a black hole must lose its electric field -- the black hole must discharge. For a small photon mass $m$, the rate of discharge can be estimated as follows. One treats the mass term $m^2\phi$, where $\phi\approx q/r$ is the electrostatic potential, as the density of the screening charge $\rho_s=-m^2\phi/(4\pi )$. One further assumes that the screening charge moves into the black hole at the speed of light. Then $\dot {q}\approx 4\pi r_g^2\rho_s=-qm^2r_g$, where $r_g$ is the Schwarzschild radius. This gives an exponential discharge, $q\propto e^{-\gamma t}$, with the decay rate $\gamma \approx m^2r_g$. In \S \ref{photon} we show that this estimate correctly describes the (intermediate asymptotic) decay of the electric field near the black hole in the limit of small photon mass.

Now consider a black hole in massive gravity. The gravitational field of the black hole is screened by a negative energy density $\rho_s\sim -m^2M_P^2$, where $m$ is the graviton mass and $M_P$ is the Planck mass. We argue in \S \ref{disappear} that this negative energy must be accreted onto the black hole. Assuming, like we did in massive electrodynamics, that the screening energy accretes onto the black hole at the speed of light, we get a decreasing black hole mass: $\dot {M}\sim -r_g^2m^2M_P^2$. Thus, the black hole loses mass and gradually disappears.

The black hole disappearance is a weird prediction of massive gravity, as it seems to make possible the following scenario: (i) there was a star in an asymptotically Minkowski space-time, (ii) the star collapses into a black hole, (iii) the black hole disappeares leaving behind just the Minkowski space-time. 

To be clear, our results are inconsequential for real astrophysical black holes. If we live in an asymptotically flat universe with a massive graviton, the graviton mass must be smaller than the Hubble constant and the disappearance time much longer than the Hubble time. In fact, the phenomenological continuity of $m\to 0$ limit is a quite satisfactory outcome. But the very possibility of disappearance (from the asymptotically Minkowski, eternal universe) seems to be a badness of the massive gravity.

What makes the above scenario impossible in General Relativity is the conservation of the asymptotically defined ADM mass. In massive gravity the ADM mass vanishes -- this is the characteristic property of the theory. And while an alternative globally conserved quantity can be defined in massive gravity (\S \ref{ADM}), global charges are not conserved in the presence of black holes. \footnote{After submitting the first version of this paper, we were informed of the work \cite{Volkov} on black holes in bimetric gravity. There, it has been argued that asymptotically flat static black holes are non-singular only when the two metrics coincide with each other, and hence, with the Schwarzschild solution of the Einstein gravity. Thinking of massive gravity as a bimetric theory in the limit where the Newton's constant $\tilde G$ of the second metric vanishes, the gravitational radius with respect to this metric is always zero, leading to the full disappearance of black holes in massive gravity. For any finite $\tilde G$, we expect the ``mass discharge'' of black holes in bimetric gravity to continue until the gravitational radii with respect to the two metrics coincide, at which point the time-dependent black hole solutions settle to the regular static solutions.}

%%%%%%%%%%%%%%%%%%%%%%%%%%%%%%%%%%%%%%%%%%%%%%
\section{\label{photon}Black hole discharge in massive electrodynamics}

In this section, we first formulate the Einstein-Proca theory and argue that charged black holes must discharge (\S\ref{EP}). We next study the decay of the electric field and define the discharge mode (\S\ref{discharge}). Finally, we find quasi-stationary time-dependent solutions in the limit of small photon mass and identify the discharge mode (\S\ref{approx_discharge}). Our massive gravity calculation will closely follow the calculation  of \S\ref{approx_discharge}.

%%%%%%%%%%%%%%%%%%%%%%%%%%%%%%%%%%%%%%%%%%%%%
\subsection{\label{EP}Einstein-Proca theory. Singularity of static charged black hole solutions}

The field of charged stars and black holes in massive electrodynamics is governed by the Proca equation 
\bea
\label{proca}
&\partial_\nu(\sqrt{-g}g^{\nu\alpha}g^{\mu\beta}F_{\alpha\beta})+m^2\sqrt{-g}A^\mu=4\pi \sqrt{-g} J^\mu,
\eea
where $F_{\mu\nu}=\partial_\mu A_\nu-\partial_\nu A_\mu$, and $J^\mu$ is the electric current. The Einstein equation is
\bea
\label{einstein}
G_{\mu\nu}=\T_{\mu\nu}+T_{\mu\nu}^{(m)},
\eea
where $\T_{\mu\nu}$ is the sress-energy tensor of matter and $T^{(m)}_{\mu\nu}$ is the stress-energy tensor of the Proca field 
\be
T^{(m)}_{\mu\nu}=T^{(0)}_{\mu\nu}+m^2(A_\mu A_\nu-\frac{1}{2}g_{\mu\nu}g^{\alpha\beta}A_\alpha A_\beta),
\ee
where $T^{(0)}_{\mu\nu}$ is the stress tensor of the Maxwell theory. For black holes and outside stars $J^\mu=0$, $\T^\mu_\nu=0$. Here, we only consider the spherically symmetric case where the metric is diagonal in $(t,r,\theta,\vphi)$ coordinates. The spherical symmetry also implies that the only non-vanishing component of $F_{\mu\nu}$ is $F_{tr}$. 

For static solutions, the only non-zero component of the Proca field is $A_t$. The $A_r$ component vanishes, because the $\mu =r$ component of equation \eqref{proca} reads
\bea
\label{dt}
\partial_t(\sqrt{-g}g^{tt}g^{rr}F_{tr})+m^2\sqrt{-g}g^{rr}A_r= 4\pi\sqrt{-g} J^r,
\eea
which forces $A_r =0$ when $\d_t=0$ and $J^r=0$. The $\mu =t$ component of \eqref{proca} then becomes
\be
\label{proca0}
\d_r(\sqrt{-g}g^{rr}g^{tt}\d_r A_t)+m^2\sqrt{-g}g^{tt}A_t= 4\pi\sqrt{-g} J^t,
\ee
and gives $A_t$. Solutions vanishing at spatial infinity have the expected Yukawa behavior at large $r$:
\be
\label{yukawa}
A_t\approx\frac{q e^{-mr}}{r}\qquad \text{when}\qquad r\gg r_g.
\ee
Defining the electric charge via the Gauss's law, this shows that localized sources are screened by the opposite charge density $-m^2 A^t/4\pi$, carried by the Proca field. 

This is how one can calculate the electric field of a charged star. But static charged black holes do not exist, as they would be singular at horizon. To demonstrate this, consider the invariant product $g^{\mu\nu}A_\mu A_\nu$, which is an observable in Proca theory. The product diverges at horizon because it reduces to $g^{tt}A_t^2$ and $A_t\neq 0$ at horizon (if $A_t$ were to vanish both at horizon and at infinity, eq.\eqref{proca0} would give $A_t(r)=0$, corresponding to an uncharged black hole).

The horizon singularity of the static solution indicates that newly formed charged black holes would get rid of their electric field hair. The corresponding time-dependent solutions will be studied in the next section. 

%%%%%%%%%%%%%%%%%%%%%%%%%%%%%%
\subsection{\label{discharge}Non-singular time-dependent black holes and the discharge mode}

For simplicity, we assume that the charge is small. Then we can use the Schwarzschild metric
\be
\label{sch0}
ds^2=(1-{r_g\over r})dt^2-(1-{r_g\over r})^{-1}dr^2-r^2d\Omega ^2.
\ee
The vacuum Proca equation for spherically symmetrical field can be transformed to an equation for the physical electric field $E=\sqrt{-F^2/2}$ (the field is, of course, radial)
\be \label{proca3}
\partial _t^2E-(1-{r_g\over r})\partial _r\left((1-{r_g\over r}){1\over r^2}\partial _r(r^2E)\right)+m^2(1-{r_g\over r})E=0,
\ee
which can be brought to another useful form:
\be \label{proca4}
(\partial _t^2-\partial _\rho^2+V)\Psi=0, 
\ee
where $\Psi=rE$, $\rho=r+r_g\ln (r/r_g-1)$ is the tortoise coordinate, and the effective potential is
\be
V=(1-{r_g\over r})({2\over r^2}-{3r_g\over r^3}+m^2).
\ee

One can now set up an initial value problem by specifying the initial $\Psi$ and $\dot \Psi $ and integrate eq.\eqref{proca4} forward in time. Since $V$ is non-singular for all $-\infty<\rho<\infty$, an initially non-singular $\Psi$ will remain so at all later times. Numerical experiments with various initial conditions confirm the anticipated decay of any initial electric field. Moreover, for small photon mass the decaying solutions asymptote to a mode that is best described by the intuitive picture of a quasi-stationary accretion of the screening charges onto the black hole: the {\it discharge mode}. 

We formally define the discharge mode by postulating the exponential in time decay of the field $\Psi (t,\rho)=e^{-\gamma t}\Psi (\rho)$. The field is in-going at horizon and decreasing at spatial infinity: 
\be \label{dmode}
(-({d\over d\rho})^2+\gamma ^2+V)\Psi=0, 
\ee
\be
{d\Psi \over d\rho}=-\gamma \Psi ,~~\rho\rightarrow -\infty,
\ee
\be
\Psi \rightarrow 0 ,~~\rho\rightarrow +\infty.
\ee
This eigenvalue problem is solved numerically in Appendix A, where we also show that our previous estimate of the discharge rate happens to be asymptotically exact: $\gamma=m^2r_g$ in the limit $mr_g\rightarrow 0$.

We emphasize that the discharge mode is not a true long-time asymptote. For $t\gg (m^2r_g)^{-1}$, the electric field will become oscillatory with an algebraic decay $t^{-5/6}\sin mt$, as shown in \cite{Konoplya}. Nor does the discharge mode uniquely describe the time evolution of the electric field outside a charged star which collapses into a black hole. The evolution will depend on how exactly the collapse occurs, and the electric field will be given by a linear superposition of the decaying initial field and the field emitted by the charges moving into the black hole. If the charges move into the black hole fast enough, it seems reasonable to assume that the radiation will be mostly beamed into the black hole. The outside field is then probably dominated by the discharge mode. But in any case, the decay of the electric field cannot occur faster than the decay of the discharge mode. So to be precise, for small $m$, the intermediate-asymptotic decay of the near-hole electric field occurs in the discharge mode.

%%%%%%%%%%%%%%%%%%%%%%%%%%%%%%%%%%%%%%%%%%%%%%%%%%%%%%%%
\subsection{\label{approx_discharge}Discharge mode as a quasi-stationary solution}

In the limit of small $mr_g$, the quasi-stationary character of the discharge mode becomes sharper. In this section we show that there exists a two-parameter family of quasi-stationary solutions characterized by the initial charge of the black hole $q$ and the discharge rate $\dot q$. The previously defined discharge mode corresponds to the unique $\dot q/q$ ratio for which the horizon is non-singular. The approximate method used in this section helps us extract analogous quasi-stationary solutions in massive gravity, where on the one hand, time-dependent simulations seem impossible, and on the other, $m r_g\ll 1$ is a reasonable assumption.

Let us return to the vacuum Einstein-Proca equations (\ref{proca},\ref{einstein}). The first observation to make is that when $m r_g\ll 1$, the expected discharge rate $\dot q \sim m^2 r_g q$ is so small that the induced time-dependence of the electric field and the metric are negligibly small as compared to the $r$-dependence. The only effect of time-dependence is to allow $A_r\neq 0$. This can be seen from the expression for the screening charge current $J_s^r\equiv -m^2A^r/4\pi$, which gives
\be
\label{qdot}
\left.\dot q = m^2\sqrt{-g}g^{rr}A_r\right|_{r=r_g}.
\ee

The second observation is that in the near-hole region we have $r\ll m^{-1}$, and as long as the fields are regular, the terms proportional to $m$ in eqs.\eqref{proca} and \eqref{einstein} are negligible. We therefore recover the time-independent Einstein-Maxwell theory, but now with an additional `gauge' condition on $A_\mu$
\bea
\label{proca1}
\partial_\mu(\sqrt{-g}g^{\mu\nu}A_\nu)=0,
\eea
which is obtained by taking the divergence of eq.\eqref{proca}. This equation happens to be independent of $m$, and in our approximation it becomes an equation for $A_r$
\be
\label{Ar}
\d_r(\sqrt{-g}g^{rr}A_r)=0.
\ee
This is just the condition of stationary flow on $J_s^\mu$. 

Assuming the Schwarzschild metric \eqref{sch0}, we solve \eqref{Ar}
\bea
A_r=\frac{c}{r(r-r_g)},
\eea
where $c$ is a new integration constant related through eq.\eqref{qdot} to the discharge rate $c=-\dot q/m^2$. As for $A_t$, we can use the Einstein-Maxwell result $A_t=q/r$. Thus, we have a two-parameter family of quasi-stationary solutions characterized by the charge $q$ and the discharge rate $\dot q$. 

Requiring the horizon to be regular uniquely fixes the discharge rate. For instance, the regularity of the norm squared $g^{\mu\nu}A_\mu A_\nu=g^{tt}A_t^2+g^{rr}A_r^2$ gives
\bea
\label{dotq}
\dot q=\pm m^2 r_g q,
\eea
which agrees with the previous calculation of the discharge mode in the limit $mr_g\to 0$. The sign is undetermined by our equations, which are invariant under the time-reversal, but physical considerations select the negative sign.\footnote{The above calculation can easily be generalized to the case where the back-reaction of charge on geometry is not negligible. The near-hole metric will then be the Reissner-Nordstr\"{o}m solution and $r_g$ in \eqref{dotq} will be replaced by $r_+$, the radius of the outer horizon.}

%%%%%%%%%%%%%%%%%%%%%%%%%%%%%%%%%%%%%%%%%%%%%%%%%%%%%%%%%%%
\section{\label{disappear}Black hole disappearance in massive gravity}

In this section, we first introduce Fierz-Pauli massive gravity, its nonlinear completions, and some of their relevant properties (\S\ref{secFP}). We next consider asymptotically flat spacetimes and argue that black holes must disappear (\S\ref{theorem}). In \S\ref{time-dep}, we find quasi-stationary solutions, and show that unlike massive electrodynamics there is no unique disappearance rate. We next examine the implications of our findings in the actual problem of star collapse (\S\ref{fate}). Finally in \S\ref{ADM}, we discuss the concept of global energy in massive gravity, and its failure in the presence of black holes. 

%%%%%%%%%%%%%%%%%%%%%%%%%%%%%%%%%%%%%%%%%
\subsection{\label{secFP}Fierz-Pauli massive gravity}

By Fierz-Pauli massive gravity (FP), we denote a class of theories described by the Einstein-Hilbert action plus the (non-linear version of) Fierz-Pauli action:
\bea
\label{S}
&S=S_{EH}+S_{FP}, &\\
&S_{EH}=-{1\over 2}\int d^4x\sqrt{-g}R,& \\
\label{FP}
&S_{FP}=m^2\int d^4x\sqrt{-g}U.&
\eea 
Here we have set $8\pi G =1$, and $U$ is a Lorentz-invariant potential defined using a flat reference metric $\eta_{ab}$. For small $h_{ab}=g_{ab}-\eta_{ab}$ it reduces to the Fierz-Pauli mass term \cite{FP}
\bea
\label{FP1}
U^{(2)}= \frac{1}{8}(h^2-h_{ab}^2).
\eea
where $h=\eta^{ab}h_{ab}$. The Lorentz-invariance of $U$ can be enforced by requiring that $U$  be a symmetric function of the eigenvalues of the matrix $H^a_b=g^{ac}\eta_{bc}$. 

The Fierz-Pauli term \eqref{FP1} is special because it is the unique quadratic expression that is linear in the perturbations of the ADM lapse function $\delta N=N-1$. So as in general relativity, $N$ remains a Lagrange multiplier and the theory describes five dynamical degrees of freedom at quadratic level. Rather obviously in the hindsight, this will no longer be the case beyond the quadratic level for a generic $U$, as Boulware and Deser pointed out \cite{BD}. This results in a sixth degree of freedom which they showed to be a ghost. This problem has been solved by de Rham, Gabadadze, and Tolley \cite{dRGT} who found a particular two-parameter family of potentials which propagates just five degrees of freedom at all orders. A concise representation of the family is \cite{Nieuwenhuizen}
\be \label{FP2}
U=\sum \lambda _a\lambda_b+\tilde{c}_2\sum \lambda _a\lambda _b\lambda _c+\tilde{c}_3\lambda _0\lambda _1\lambda _2\lambda _3,
\ee
where the sums are over all all-distinct pairs and triples of indices, and $\lambda _a$ are the four eigenvalues of the matrix
\be
\label{1-H}
\delta ^a_b - \sqrt{ H^a_b }.
\ee
It is easy to understand why this two-parameter family, which we henceforth refer to as FP2, is special. Restricting to metrics with zero shift vector, the lapse appears only in $\lambda_0=1-N^{-1}$. Apparently, the expression \eqref{FP2} is the only symmetric combination of $\lambda_a$, such that $\sqrt{-g}U_{FP2}$ has zero cosmological constant, and is linear in $N$. The full proof of the absence of Boulware-Deser ghost is given by Hassan and Rosen \cite{Hassan} (see also \cite{proof} for a different approach).\footnote{Evidently, having the right number of degrees of freedom does not protect FP2 against other pathologies such as strong coupling, instability, or superluminality of fluctuations when the background metric deviates from Minkowski (see, e.g., \cite{deRham_dS,Gruzinov_suplum,deRham_suplum,Deser}). The asymptotically flat solutions considered here are not expected to be exceptional \cite{Berezhiani_fluctuation}.}

As formulated above, the FP theory is not generally covariant. However, it can be made so by introducing 4 non-canonical scalar fields \cite{AGS,Dubovsky,Chamseddine} to write $H^a_b$ as 
\be
\label{H}
H^a_b=\eta _{bc}g^{\mu \nu}\partial _\mu\phi ^a\partial _\nu\phi ^c.
\ee
In the so-called unitary gauge, one uses these scalar fields as coordinates, $X^a=\phi^a$, and recovers the original formulation of the FP theory. In the unitary gauge, the theory is still invariant under simultaneous reparametrization of $g_{ab}$ and $\eta_{ab}$. For our purposes -- studying spherically symmetrical stars and black holes -- we transform to spherical coordinates. Then \comment{In the unitary gauge, coordinate transformations are no longer a symmetry of $S_{FP}$ but change the `coupling constants' $\eta_{ab}$. When solving spherically symmetric problems, we can use these transformations to simultaneously rotate the physical and the reference metrics into spherical coordinates, where}
\bea
\label{eta}
\eta_{ab}dX^a dX^b=dT^2-dR^2-R^2d\Omega^2\,.
\eea
If the space-time is spherically symmetric, the most general metric in this coordinate system is parametrized by four functions of $T$ and $R$:
\bea
\label{g}
ds^2=C dT^2-2 D dT dR -A dR^2- B R^2 d\Omega^2\,.
\eea

{\em A comment on our notation:} in the following we will frequently switch to the time variable $t$ as measured by asymptotic observers, and the circumference defined radius $r$ (the Schwarzschild variables). However, the numerical indices $(0,1,2,3)$ are exclusively used to denote unitary-gauge variables in the spherical coordinates in the order $(T,R,\theta,\varphi)$.

%%%%%%%%%%%%%%%%%%%%%%%%%%%%%%%%%%%%%%%%%%%%%%%%%%%%%%%%%%%%%
\subsection{\label{theorem}Asymptotically flat spacetimes and the inevitable singularity of static black holes}

As in massive electrodynamics, it is natural to first look for static star and black hole solutions in massive gravity. The procedure is briefly outlined in appendix \ref{static}, and it was fully pursued in \cite{stars} with the conclusion that there exist acceptable star solutions but black holes are generically singular at horizon. However, again as in the case of massive electrodynamics, it is easy to exhibit the horizon singularity of static black holes without knowing the explicit solution. The horizon singularity now indicates that a star collapsing into a black hole must get rid of its very gravitational field and hence disappear.\footnote{Sergei Dubovsky pointed out that the horizon singularity of our static black hole solution \cite{stars} might signal accrertion, like it does for fluids. If one tries to find a static fluid surrounding a black hole, one gets a singularity at horizon; in reality fluids are accreted.}

The horizon singularity can be shown as follows. Consider the most general static spherically symmetric vacuum solution of massive gravity. The unitary-gauge metric is given by \eqref{g}. Requiring the space-time to be asymptotically flat then forces the metric to be diagonal \cite{stars}, i.e. $g_{01}=D=0$:

\vspace{-0.3 cm}

\begin{quote}

{\it First note that the Ricci tensor of a time-independent metric of the form \eqref{g} satisfies the identity \cite{Salam} 
\be
\label{gR}
g_{01}R_{00}-g_{00}R_{01}=0.
\ee
This imposes the following purely algebraic constraint on the unitary-gauge metric components via the vacuum massive gravity equations:
\bea
\label{constraint}
g_{01}T_{00}-g_{00}T_{01}=0\,,
\eea
where $T_{ab}$ is the stress-energy tensor of $S_{FP}$. 

On the other hand, for any potential $U$ that is a symmetric function of the eigenvalues of the matrix $H^a_b=g^{ac}\eta_{cb}$, one can verify that $T_{01}=\kappa g_{01}$, where the proportionality coefficient $\kappa$ is a non-singular function at $g_{01}=0$ (see footnote \ref{kappa}). It follows that eq. \eqref{constraint} divides the solutions into two branches

(i) $g_{01}=0$,

(ii) $T_{00}=\kappa g_{00}$.

Now consider large radii $R\to \infty$, where the metric perturbations $h_{ab}= g_{ab}-\eta_{ab}$ are small by the asymptotic flatness assumption. For any non-linear completion of the Fierz-Pauli mass term $h_{ab}^2-h^2$, we then have
\bea
T_{01}=\frac{1}{2}m^2h_{01}(1+{\cal O}(h))\,,
\eea
giving $\kappa =m^2/2$ at large $R$. This excludes the branch (ii) as it leads to a finite asymptotic value for $T_{00}$. This branch is where the Schwarzschild-de Sitter solutions of \cite{Salam,Koyama} are realized (see also \cite{Berezhiani} in that context).
}
\end{quote}

On the other hand, whenever the unitary-gauge metric is diagonal, horizons will be physically singular in massive gravity \cite{Deffayet}. This is because the inverse unitary-gauge metric components are scalar quantities in massive gravity: $g^{ab}=g^{\mu\nu}\d_\mu\phi^a \d_\nu \phi^b$. When $g^{ab}$ is diagonal, its $\{00\}$ component will be singular at horizon and this singularity will be reflected in the action and the stress-energy tensor via $\lambda_0=1-\sqrt{g^{00}}$. To have finite $g^{ab}$ at horizon, one necessarily needs $g_{01}\neq 0$ which, by the above arguments, leads to time-dependence and non-zero energy flux $T^1_0$. \footnote{\label{t}Another way to see why $g_{01}\neq 0$ is a necessary condition for regularity of the horizon is to realize that when $g_{01}=0$, the unitary-gauge time variable $T=\phi^0$ coincides with the proper time of the asymptotic observers $t$. But $t$ is a singular variable at horizon and therefore $\phi^0=t$ is singular \cite{Jacobson}. This inevitable singularity relies just on the existence of a horizon, but if the Schwarzschild geometry is indeed recovered at short distances \`{a} la Vainshtein, we know how exactly $t$ diverges at horizon: in terms of the advanced time $v$ and circumference defined radius $r$ which are regular variables
\be
t=v-\rho=v-[r+r_g\ln(r/r_g-1)].
\ee}

In view of black hole discharge in massive electrodynamics, the time-dependence of black holes in massive gravity is in fact naturally expected. The linearized field of a point source is known (and shown in \S \ref{lin}) to exhibit the Yukawa decay $h_{ab}\propto \exp(-mr)$ in massive gravity. It follows that the invariant ADM mass of any localized system is zero. Therefore, the mass term, taken to the right hand side of the Einstein equation, can be thought of as the stress tensor of screening matter with negative energy that surrounds and degravitates gravitational sources. When a black hole forms, the screening matter flows inside and diminishes the black hole mass.\footnote{The classical instability of black holes in massive gravity has been conjectured before by Gia Dvali \cite{Dvali}.}

The rate of this process is estimated by requiring the near horizon energy density $T^0_0\sim -m^2$ accrete with the speed of light through an area of order $r_g^2$:
\bea
\label{rgdot}
\dot r_g\sim -m^2r_g^2,
\eea
where for a mass $M$ black hole $r_g=M/4\pi$ in our unites. Note that when $mr_g\ll 1$ the associated scale $\tau=1/m^2 r_g$ is much longer than other length scales in the problem. We will use this fact to find approximate time-dependent solutions in the next section.

Assuming that the time-dependence is mild, of order \eqref{rgdot}, we can estimate the resulting value of the off-diagonal metric component $g_{01}$. Note first that the identity \eqref{gR} is violated in the time-dependent case by terms of order 
\be
g_{01}R_{00}-g_{00}R_{01}\sim \partial_0\partial_1 g_{00}\sim \dot r_g/r^2\sim m^2 r_g^2/r^2,
\ee
where we used the estimate \eqref{rgdot} in the last step. Next replace $R_{ab}$ by $T_{ab}$ via the Einstein equation, as we did to find the constraint \eqref{constraint}, and use the same arguments to find
\be
g_{01}(T_{00}-\kappa g_{00})\sim m^2 r_g^2/r^2.
\ee
Since $g_{01}=0$ as $r\to \infty$ and $\kappa \sim m^2$, we get $g_{01}\sim r_g^2/r^2$. Thus, we expect the diagonal static solution to be a good approximate solution everywhere except very close to the horizon where $g_{01}$ becomes of order unity and makes the inverse metric $g^{ab}$ finite. This expectation will be confirmed by the calculation of the next section.

%%%%%%%%%%%%%%%%%%%%%%%%%%%%%%%%%%%%%%%%%%%%%%%%%%%%%%%%%%%%
\subsection{\label{time-dep}Non-singular time-dependent black holes}

It is relatively easy to find exact time-dependent solutions in the linear regime, but this is not the case at non-linear level. We therefore break the problem into three parts: first, we find the exact linear solutions and show that they are characterized by two parameters, the mass $r_g$ and the disappearance rate $\dot r_g$. Then we use the $mr_g\ll 1$ approximation to find quasi-stationary near-hole solutions, again parametrized by the mass and rate. Finally, we match the linear and the near-hole solutions. We will see that unlike massive electrodynamics where all but a unique ratio $\dot q/q$ lead to a singularity at the horizon, here all quasi-stationary solutions are regular. The analogous case of the accretion of superluminal fluids and possible explanations of this result are discussed in the following section \S\ref{fate}. 

%%%%%%%%%%%%%%%%%%%%%%%%%%%%%%%%%%%%%%%%%%%%%%%%%%%%%%%
\subsubsection{\label{lin}Linear solutions}

To study the linearized system, we parametrize the metric as
\bea
ds^2= (1+c)dT^2-2d \;dT dR -(1+a) dR^2 - (1+b) R^2 d\Omega ^2,
\eea
where $a,b,c,d$ are infinitesimal functions of $T$ and $R$. The linearized Einstein equation in Cartesian coordinates reads 
\bea
\label{elin}
\Box h_{ab}-\d_a\d_c h^c_b-\d_b\d_c h^c_a+\d_a\d_b h -\eta_{ab}\Box h +\eta_{ab}\d_c\d_d h^{cd}=m^2(\eta_{ab}h-h_{ab}),
\eea
and its divergence yields the condition $\partial_a h^a_b=\d_b h$. Since the equations are first order in perturbations $h_{ab}$, we can replace in them $(T,R)$ with $(t,r)$ which coincide at zeroth order. Moreover, the solutions can be expanded in the exponential basis
\bea
a,b,c,d \propto e^{-\lambda m t}.
\eea
The system of equations can then be brought into the form
\bea
a' r+2a-2b=\frac{\lambda^2}{1-2\lambda^2}m^2 r^2 (a - 2\lambda^2 b),\\
b' r+b -a = \frac{1}{2(1-2\lambda^2)}m^2 r^2 (a - 2\lambda^2 b),
\eea
where $'\equiv d/dr$, and $c$ and $d$ are given in terms of $a,b$ by
\bea
c'r &=& 2(1+\lambda^2)(a-b-b'r),\\
\label{d}
d&=&\frac{2\lambda}{m r}(a-b-b'r).
\eea
Thus, for any fixed rate $|\lambda|<1/\sqrt{2}$ there exists a one-parameter family of decaying as a function of $r$ solutions. The parameter characterizes the mass of the gravitating body. 

Note that in agreement with the arguments of the previous section, $g_{01}$ vanishes in the static limit $\lambda \to 0$. For an accretion rate of order \eqref{rgdot}, we have $\lambda \equiv \alpha m r_g \ll 1$ (where we introduced $\alpha$ as a so far undetermined order-unity parameter characterizing the rate $\dot r_g$). The solution can therefore be approximated by the $\lambda=0$ vDVZ solution \cite{vDVZ}:
\bea
c = -\frac{4r_g}{3r} e^{-mr},\quad a=-\frac{4r_g}{3m^2r^3} e^{-mr}(1+mr),\quad b=\frac{2r_g}{3m^2r^3} e^{-mr}(1+mr+m^2r^2),
\eea
except that $r_g$ adiabatically changes with time as $r_g(t)\simeq r_g(0) \exp(-\alpha m^2 r_g t)$, and
\bea
\label{d1}
d = \frac{4\alpha r_g^2}{3r^2} e^{-mr}(1+mr).
\eea
We see that, apart from allowing small non-zero $g_{01}$, the time-dependence is inconsequential in the long-distance linear field.\footnote{We also see that in the context of massive gravity the no-hair theorem of \cite{Hui_nohair} and in general the singularity of static black holes cannot be interpreted as a dramatic difference between the field of stars and black holes. This can be understood from the fact that the linear static field of stars and black holes is uniquely determined by the mass $r_g$. To lose its hair, the black hole must actually lose its entire mass, which is done only very slowly. In particular, we think that contrary to the claims of \cite{Hui_galaxy}, the response of stars and black holes of the same mass to the field of distant objects is almost indistinguishable.} In FP2, this linear solution is valid for $r\gg r_V=(r_g/m^2)^{1/3}$.

%%%%%%%%%%%%%%%%%%%%%%%%%%%%%%%%%%%%%%%%%%%%%%%%%%%%%%%%%%%%
\subsubsection{\label{near}Near-hole solutions}

Similar to massive electrodynamics, we can develop an approximate method to find the short distance ($r\ll r_V$) quasi-stationary solutions in the limit $mr_g\ll 1$. This approximation is in fact a generalization of the Vainshtein's original idea \cite{Vainshtein}, which we now review and extend to our case.

Aiming for a static solution, Vainshtein started from a diagonal metric ansatz
\bea
\label{unit}
ds^2=C dT^2-A dR^2 -BR^2 d\Omega^2,
\eea
where $A,B,C$ are functions of $R$, and are determined using the Einstein equations
\bea
\label{einstein1}
G^\mu_\nu=T^\mu_\nu.
\eea
He noticed that $T^\mu_\nu$ is proportional to $m^2$ and therefore can be ignored at short radii, as long as the metric coefficients remain finite. Therefore, the mass-less Einstein theory of gravity is recovered in this limit, implying that the metric \eqref{unit} must be a reparametrization of the Schwarzschild metric
\bea
\label{sch}
ds^2=(1-\frac{r_g}{r})dt^2-(1-\frac{r_g}{r})^{-1}dr^2-r^2 d\Omega^2.
\eea
The reparametrization is determined from the covariant divergence of eq.\eqref{einstein1}:
\be
\label{dT1}
\nabla_\mu T^\mu_\nu=0,
\ee
which can be thought of as a gauge condition on the metric \eqref{unit}. Note that $m$ drops out of this equation, so it is a non-trivial constraint even in the zero-graviton-mass limit. 

Requirement of asymptotic flatness ($g_{ab}=\eta_{ab}$ as $R\to \infty$) then fixes $T=t$, and the only non-trivial component of eq.\eqref{dT1} (the $\nu=1$ component) serves as an equation for $R(r)$, in terms of which $A,B,C$ are given by
\bea
C=1-\frac{r_g}{r},\quad A = 1/(C{R'}^{2}),\quad B = r^2/R^2.
\eea
Vainshtein realized that while there is no such reparametrization of the Schwarzschild metric at the linearized level (the well-known vDVZ discontinuity), linearization becomes inadequate at $r\sim r_V$. He showed that at the non-linear level there exists a finite solution valid for $r_g\ll r\ll r_V$, which a posteriori justifies neglecting $T^\mu_\nu$ from the Einstein equation. One still needs to check whether this solution matches the vDVZ solution for $r\gg r_V$, where the mass term cannot be ignored anymore. This is a non-trivial check, but there exists a sub-family of FP2 for which the answer is positive \cite{stars,Chkareuli} (see \cite{Babichev} for earlier works, and \cite{Berezhiani_restricted} for a related discussion). 

Returning to the time-dependent problem, we expect that, as in massive electrodynamics where for small $mr_g$ the only relevant effect of the time-dependence was to excite $A_r$, here the time-dependence excites $g_{01}$. Once this is taken into account by using the metric ansatz 
\bea
\label{u_metric}
ds^2=C dT^2-2D dT dR -A dR^2 - B R^2 d\Omega^2,
\eea
we can ignore time derivatives (quasi-stationary approximation) and repeat Vainshtein's procedure. Neglecting the stress tensor of the FP action from the Einstein equation implies that the metric is a special reparametrization of the Schwarzschild metric which satisfies the gauge condition $\nabla_\mu T^\mu_\nu=0$.

In the quasi-stationary approximation, the metric \eqref{u_metric} can be diagonalized and expressed in terms of $t$ and $r$ by the coordinate transformations
\bea
R&=&B^{-1/2}r,\\
\label{T}
T&=&t+\int\frac{D}{C}\gamma dr,
\eea
where $\gamma\equiv dR/dr=B^{-1/2}(1-B'r/2B)$. In this coordinate system the metric looks like
\bea 
ds^2 = C dt^2 -\gamma^2(A+D^2/C)dr^2-r^2d\Omega^2.
\eea
Requiring this to match the Schwarzschild metric \eqref{sch} gives
\bea
\label{CA}
C=1-r_g/r,\qquad A=C^{-1}(\gamma^{-2}-D^2).
\eea
Hence, there are two unknowns $D$ and $B$, characterizing the gauge transformation from the unitary frame to the Schwarzschild frame. The task is to solve for them using the $\nu=0,1$ components of $\nabla_\mu T^\mu_\nu=0$ on a fixed Schwarzschild geometry, and subject to appropriate boundary conditions. The resulting quasi-stationary solutions are expected to be parametrized by the mass $r_g$ and the disappearance rate $\dot r_g=\alpha m^2 r_g^2$ (which will be the integration constant of the $\nu =0$ equation). One must then ask for what values of $\alpha$ the solution is regular at horizon.\footnote{\label{phi0}In the language of footnote \ref{t}, one seeks solutions on which $\phi^0$ interpolates between $\phi^0=t$ at $r=\infty$ and $\phi^0=t+\rho$ at $r=r_g$ \cite{Jacobson}.}

Note that the above procedure is equivalent to solving the equations of motion for the scalar fields $\phi^a$ on a fixed Schwarzschild background (as one usually treats accretion problems \cite{Bondi}). By spherical symmetry $\phi^i= R n^i = B^{-1/2}x^i$, and the stationary approximation corresponds to looking for solutions of the form $\phi^0 = t+\varphi(r)$ [which is related to $B$ and $D$ by eq.\eqref{T}]. This is why there are only two non-trivial equations; they are related to the stress-energy conservation via the Bianchi identity
\be
\nabla_\nu T_\mu^\nu=-\frac{1}{\sqrt{-g}}\partial_\mu\phi^a\frac{\delta S}{\delta \phi^a},
\ee
which is particularly simple in the unitary gauge where $\d_b \phi^a=\delta^a_b$. Neglecting the back-reaction on geometry is justified for small graviton mass if we find a regular solution.

The details of the calculation for FP2 is given in appendix \ref{app-near}. In summary, we obtain an algebraic equation for $\l_2=1-B^{-1/2}$:
\bea
\label{B0}
\beta (1-\frac{3r_g}{4r})+(c_2+c_3\l_2)(1+(1-\l_2)C)=\left(C+\frac{\alpha^2r_g^4}{\beta^2 r^4}\right)^{1/2}(1+c_2\l_2+2(c_2+c_3\l_2)),
\eea
where $c_{2,3}$ are related to the parameters of FP2 action \eqref{FP2}, and $\beta\equiv 1+2c_2\l_2+c_3\l_2^2$ is a positive function when $c_2^2\leq c_3$, the range where stable numerical time-independent solutions exist. Given $\alpha$ and $r_g$, eq.\eqref{B0} is solved for $\l_2$, using which $D$ is given by \footnote{The uniqueness of the solution once $r_g$ and $\dot r_g$ are fixed is a sign of the absence of Boulware-Deser ghost in FP2. Among the five degrees of freedom of a massive graviton only the scalar mode is dynamical in the spherically symmetric problem, and can participate in the accretion (analogous to the longitudinal polarization of the massive photon and the sound mode of fluids). After the large-$r$ asymptotic condition is fixed by the mass $r_g$ (charge $q$ for photon, asymptotic density for fluids), only one additional integration constant is needed to fully parametrize the stationary accretion. In generic FP, on the other hand, the analogue of equation \eqref{B0} for $B$ is of the second order, which requires two extra integration constants. These correspond to the emergence of a second scalar degree of freedom, which is the Boulware-Deser ghost.}
\be
\label{D0}
D=\frac{\alpha r_g^2}{\beta r^2}(\g^{-1}+C)\left(C+\frac{\alpha^2r_g^4}{\beta^2 r^4}\right)^{-1/2}.
\ee

Now we can estimate how much the time-dependent solutions deviate from the static solution. Assuming $\alpha \sim 1$, equation \eqref{D0} gives a small off-diagonal metric component at $r\gg r_g$:
\be
D \simeq \frac{\alpha r_g^2 (1+\g^{-1})}{\beta r^2},
\ee
and eq.\eqref{B0} reduces to its static $\alpha=0$ limit, giving $\lambda_2=1/\sqrt{c_3}$. Therefore, as anticipated in \S\ref{theorem} the effects of time-dependence are felt merely within a distant of a few $r_g$ from the horizon, beyond which the static solution of \cite{stars} remains accurate. In particular, matching to the linearized solution (studied in appendix \ref{trans}) does not impose any constraint. This should not be very surprising in view of the fluid accretion problem. The critical radius that determines the accretion rate for fluids is close to the horizon for relativistic fluids. This is also what one expects from the example of massive electrodynamics where regularity of the horizon itself determines the accretion rate.

What is the actual rate of accretion $\dot r_g$? In massive electrodynamics, the discharge solution has been selected from the two-parametric $(q,\dot q)$ family of quasi-stationary solutions by the horizon non-singularity. If massive gravity was fully analogous to electrodynamics, the rate $\dot r_g$ would also be uniquely determined by the horizon non-singularity. But this does not actually happen. For $r\to r_g$ (and $C\to0$), eq.\eqref{D0} gives
\bea
D=\epsilon\g^{-1}+\epsilon C\left(1-\frac{\beta^2 r^4}{2\alpha^2 r_g^4 \g}\right),\qquad \epsilon = \frac{\alpha}{|\alpha|},
\eea
and upon substitution in \eqref{T} gives $\phi^0=T=t+\epsilon\rho+\mathcal{O}(\alpha^{-2})$, which is regular at black hole horizon for any positive $\alpha$ (see footnote \ref{phi0}).\footnote{If one only requires the regularity of $g^{ab}$, then negative values of $\alpha$ are also admissible. In that case, the black hole excretes negative energy `aether' and grows.} So in contrast to massive electrodynamics, any non-zero rate seems to result in a regular solution in massive gravity. This behavior, though peculiar, is not unprecedented. It is known \cite{Babichev_accretion} (and reviewed in appendix \ref{fluid}) that the accretion rate of fluids with superluminal speed of sound onto black holes is not unique. In the next section, we examine the implications of this finding for the actual process of star collapse, where we argue that the rate is determined by the history of collapse.

%%%%%%%%%%%%%%%%%%%%%%%%%%%%%%%%%%%%%%%%%%%%%%%%%%%%%%%%%%%%
\subsection{\label{fate}The fate of collapsing stars}

In classical physics, an outside observer never sees a fully formed black hole -- collapsing stars are stuck forever at their gravitational radius with redshift increasing at their surface. In this section, we study black hole accretion from this perspective which, besides being more realistic, provides a useful thought laboratory to examine our analytic results. We first consider the accretion of fluids onto collapsing stars. We then give a related interpretation of black hole discharge in massive electrodynamics, and finally, turn to the problem of collapsing stars in massive gravity.

%%%%%%%%%%%%%%%%%%%%%%%%%%%%%%%%%%%%
\subsubsection{Fluids}

Ideal fluids are characterized by their density, pressure, and velocity fields, respectively $\vep,p$, and $u^\mu$, and a certain equation of state that relates $\vep$ and $p$. The  fluid stress-energy tensor is 
\be
\T_{\mu\nu}=(\vep+p)u_\mu u_\nu- p g_{\mu\nu}.
\ee
Consider a spherically symmetric static solution on the Schwarzschild metric (we ignore the fluid back-reaction on geometry). The $\nu =1$ component of the stress-energy conservation, $\nabla_\mu \T^\mu_\nu=0$, gives 
\be
p'=\frac{-r_g}{2r(r-r_g)}(\vep+p).
\ee
Assuming a simple $p=c_s^2 \vep$ equation of state, we get the density field 
\be
\label{density}
\vep = \vep_0 (1-\frac{r_g}{r})^{-(1+c_s^{-2})/2},
\ee
where $\vep_0$ is the density at spatial infinity. We see that for subluminal and superluminal fluids alike the density and pressure diverge at $r=r_g$, reflecting the fact that an infinite force is needed to hold a test particle at rest above the horizon. 

%%%%%%%%%%%%%%%%%%%%%%%%%%%%%%%%%%%%%%%%%%%%%%%%%%%%%%%%%%%
\begin{figure}[t]
\begin{center}
\includegraphics[width=8.cm]{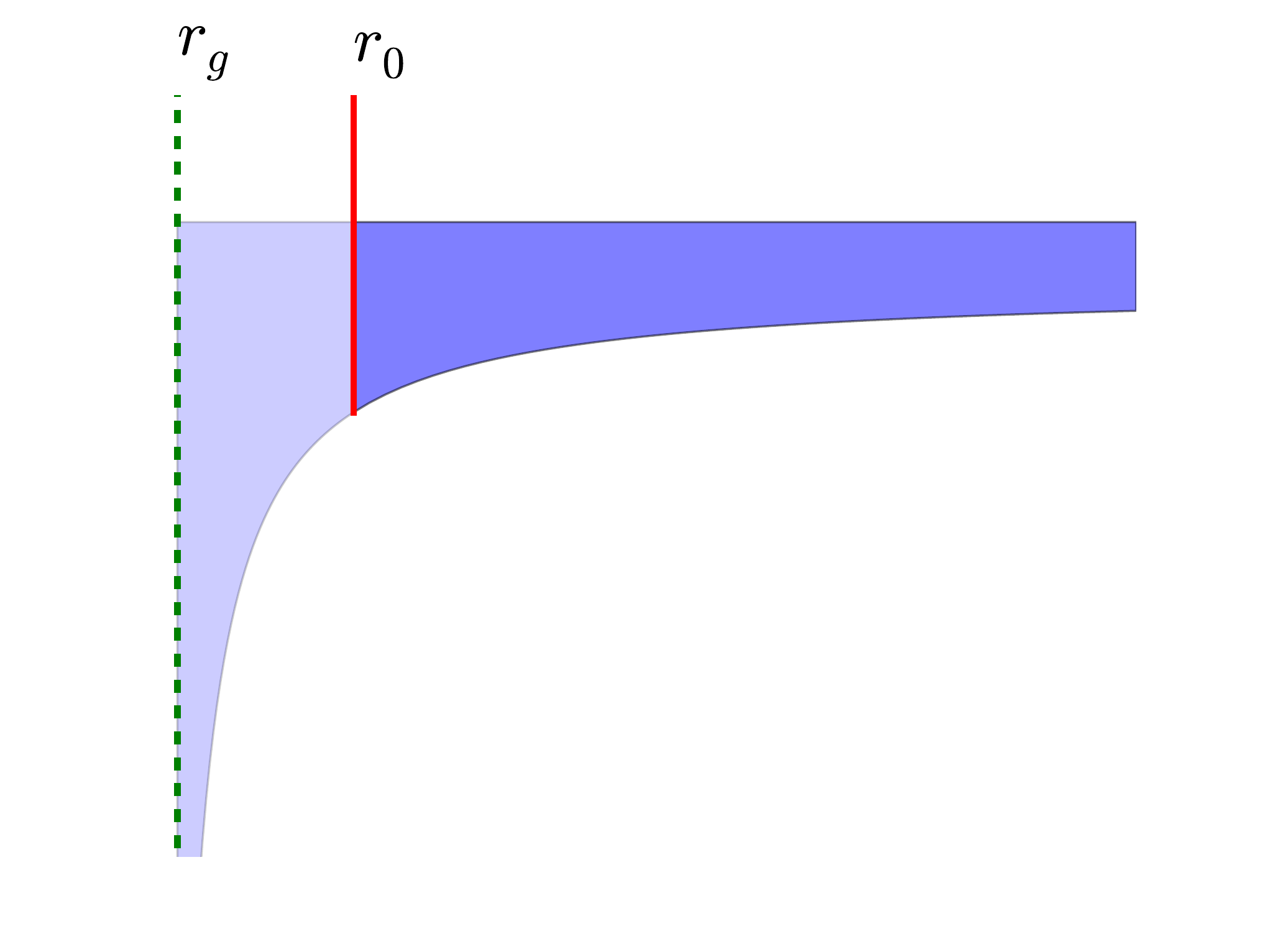}
\caption[]{ \small{The density of static fluids (represented by the depth) always diverges at horizon. However, the total amount of fluid mass that can be stored at rest between the star surface and the horizon is infinite for $c_s^2\leq 1$, while finite for $c_s^2>1$.} }

\label{stat}

\end{center}
\end{figure}
%%%%%%%%%%%%%%%%%%%%%%%%%%%%%%%%%%%%%%%%%%%%%%%%%%%%%%%%%%%%%%%%

However the two cases,  subluminal and superluminal, are known to be very different when one consideres the black hole accretion. The rate of accretion is uniquely determined by the critical Bondi solution in the subluminal case, $c_s^2\leq 1$, because accreting solutions with slower rates are singular at horizon, while solutions with faster rates do not exist. On the other hand, in the superluminal case, when $c_s^2>1$, the critical accretion rate does not exist and all accreting solutions are regular at horizon (see appendix \ref{fluid}).

This difference is ultimately due to the different amounts of {\em static} fluid mass which can be stored near the black hole horizon. On Schwarzschild metric, there is a well-defined notion of fluid mass since the $\nu=0$ component of the stress-energy conservation becomes $\d_\mu (\sqrt{-g}\T^\mu_0)=0$ and implies the conservation of
\be 
\label{M}
M=\int d^3x \sqrt{-g}\T^0_0.
\ee
At spatial infinity where the fluid is at rest and the metric is flat this reduces to $\int d^3\mbf{r}\vep$. Hence, $M$ is the fluid energy as measured by asymptotic observers. Now consider a high redshift star of radius $r_0$ as a model for the black hole. Using the static solution \eqref{density}, it is easy to see that when $c_s^2<1$ the fluid mass near the star surface diverges as $r_0\to r_g$, while it remains finite for $c_s^2>1$ (figure \ref{stat}).

Imagine next initiating the fluid accretion by contracting the star (classically, $r_0$ will pass the horizon only at $t=\infty$, no matter how fast the contraction takes place, so the accretion problem can be fully addressed in this framework). When $c_s^2<1$, the above-mentioned divergence of near-horizon fluid mass makes it possible to sustain an infinitesimal rate of steady accretion by adiabatically contracting the star. The pressure and density in this regime are then approximately given by the singular static solution. This explains the singularity of those sub-critical accreting solutions. In contrast, because of the finiteness of near-horizon $M$ in the $c_s^2>1$ case, it is impossible to sustain any steady accretion by adiabatic contraction of the star. Consequently, the accreting solutions of superluminal fluids can never be approximated by the singular static solution near the horizon.

Without an explicit calculation of fluid accretion we cannot proceed much further. So in the following, besides filling in some of the details of the above picture, we use the results of appendix \ref{fluid} to provide a more comprehensive description. 

Let us define $x=r-r_g$ in terms of which the $t-r$ part of the near-horizon Schwarzschild geometry is given by the Rindler metric:
\be
ds^2=r_g^{-1}x dt^2-r_g x^{-1}dx^2.
\ee
For a radially moving time-like curve near the horizon, we have
\be
\dot x^2=r_g^{-2}x^2(1-r_g^{-1}x e^{-2}),
\ee
where the overdot denotes $t$-derivative, and $e\equiv u_0$ is the energy per unit rest mass as measured from infinity. $e$ is constant for free-falling observers, but it vanishes as $e=(x/r_g)^{1/2}$ for static ones. Thus, the maximal value of $|\dot x|$ near the horizon is $|\dot x|_{\rm max}=x/r_g$. 

As a function of the position of the star surface $x_0=r_0-r_g$, the mass of static fluid with density \eqref{density} that resides near the surface scales as 
\be
x_0^{(1-c_s^{-2})/2}.
\ee
When $c_s^2 < 1$, this diverges in the limit $x_0\to 0$ and, as explained above, infinitesimal rates of steady accretion can be obtained by adiabatically decreasing $x_0$:
\be
\label{Mdot}
\dot M= -4\pi r_g^2 \vep \dot x_0,
\ee
with $\vep$ given by \eqref{density}. As $|\dot x_0|$ is increased to the point that this rate exceeds the critical Bondi rate, the fluid dynamics decouples from the motion of the surface and settles to the regular Bondi solution. 

On the other hand, when $c_s^2>1$ substituting the maximal value $|\dot x_0|_{\rm max}=x_0/r_g$ in eq. \eqref{Mdot} gives zero as $x_0\to 0$. Thus, similar adiabatic contractions do not correspond to any steady accretion for superluminal fluids. This result may seem rather counterintuitive as it suggests that steady accretion is impossible when $c_s^2>1$. This is of course not the case. Consider again the near horizon contribution to $M$ for a possibly moving fluid:
\be
M(x_0)\sim 4\pi r_g^2\int_{x_0} dx[(\vep+p)r_g e^2x^{-1}-p],
\ee
where the expression in the square brackets is $\T^0_0$ for a fluid with $u_0=e$. Substituting the infinitely redshifted $e=(x/r_g)^{1/2}$ for a static fluid, we see that the density of superluminal fluids is not singular enough and the integral converges as $x_0\to 0$. However, if the same fluid is allowed to fall in, with $e\neq 0$ at $x= 0$, even with a finite $\vep$ and $p$ an infinite amount of mass can be deposited near the horizon. The flux
\be
\T^1_0=(\vep +p)r_g e^2 x^{-1} \dot x,
\ee
would then be non-zero. The explicit solution of appendix \ref{fluid} shows that any non-zero accretion rate of superluminal fluids can indeed be obtained by a specific choice of $e$ at horizon. So different free-fall ($e=\rm{const.}$) motions of the star surface at horizon lead to different rates of accretion, implying that unlike subluminal fluids the accretion of a superluminal fluid never decouples from the star collapse. Moreover, since even for infinitesimal accretion rates the fluid is free-falling at horizon, it is expected that the pressure remains finite there. 

Given that at any finite $t$ all of the accreted matter is still above the horizon, one may wonder what happens if the contraction of the star is paused after a long period $\Delta t$ of steady accretion. Clearly the static solution must be recovered since the star surface is at a finite redshift. The finiteness of the near horizon static fluid mass when $c_s^2>1$ then implies that, for large enough $\Delta t$, the star must excrete most of the fluid that has been accumulated near the surface back to larger radii. 

Finally, when $c_s^2=1$ the total mass of static fluid adjacent to the surface diverges logarithmically with $x_0$. Therefore, in the adiabatic approximation \eqref{Mdot} even infinitesimal rates require very fast motion ($|\dot x_0|\propto x_0$, although with $e\propto x^{1/2}$) which makes this approximation unreliable. A more detailed analysis is needed to show the singularity of sub-critical accreting solutions in this case.

%%%%%%%%%%%%%%%%%%%%%%%%%%%%%%%%%%%%%%%
\subsubsection{Massive electrodynamics}

Now let us interpret the black hole discharge in massive electrodynamics by considering a charged shell of radius $r_0$ approaching its gravitational radius $r_g$. We first need to determine the relation between the asymptotic field $A_t=q e^{-mr}/r$ and the actual charge $Q$ on the shell as a function of $r_0$. This can be obtained from the discontinuity of the electric field at the surface in the following way (a more detailed analysis can be found in \cite{Dolgov}):

\vspace{-0.3 cm}

\begin{quote}

{\it Ignoring back-reaction on geometry, and substituting $A_t=q \chi/r$ in the source-free Proca equation, we get for $r>r_0$
\be
\chi''=m^2(1-\frac{r_g}{r})^{-1}\chi.
\ee
For small $mr_g$, the decaying solution at infinity $\chi = e^{-mr}$ goes into 
\be
\chi=1+m^2r_g x\ln x
\ee
near the horizon. On the other hand, inside the shell $r<r_0$, where $g_{rr}=-1$ and $g_{tt}\simeq x_0/r_g = {\rm const.}$, we have
\be
\chi''=m^2\chi.
\ee
Imposing regularity condition at the center and matching to the outside solution we get for $r< r_0$  
\be
A_t= \frac{q}{r_0}\frac{\sinh mr}{mr}.
\ee
The charge on the shell is then given by the discontinuity of proper electric field at $r_0$
\be
\label{sigma}
\left. Q = \sqrt{-g}g^{rr}g^{tt}F_{rt}\right|_{r_0^-}^{r_0^+}
\simeq q m^2r_g^2  \left[\ln (r_g/x_0)+\frac{1}{3}(r_g/x_0)^{1/2}\right],
\ee
where we have kept both the contributions from outside and inside although the former is sub-dominant in the $x_0\to 0$ limit.}
\end{quote}

To have a fixed asymptotic field (characterized by $q$), it is seen from eq.\eqref{sigma} that the total charge on the shell $Q$ must be increased indefinitely as the surface redshift is taken to infinity. This is due to the divergence of the total amount of screening charge both inside and outside of the shell, respectively as $x_0^{-1/2}$ and $\ln x_0$. In the realistic problem, as a star of fixed charge $Q$ contracts to smaller radii, its asymptotic field diminishes since the surrounding screening charges accrete into the now deeper potential well around and inside the star. 

As in the case of subluminal fluids the divergence of total static charge in the limit $x_0\to 0$ allows us to reproduce small rates of accretion by slowly contracting the shell:
\be
\dot q = -  \frac{\d Q}{\d x_0}\dot x_0\simeq \frac{1}{6}q m^2 r_g^{5/2} x_0^{-3/2}\dot x_0.
\ee
However, here the screening charges do penetrate inside the shell and the slow motion of the surface does not imply that of the charges. In fact, the logarithmic divergence of charge density outside the shell makes this case analogous to fluids with $c_s=1$, and a detailed analysis is needed to show why discharge rates of less than $m^2 r_g$ lead to a singular horizon. As the shell collapses faster, the outside field eventually decouples from the motion of the shell and decays via the discharge mode.

%%%%%%%%%%%%%%%%%%%%%%%%%%%%%%%%%%%%%%%%%%
\subsubsection{Massive gravity}

For static black holes in massive gravity, we show in appendix \ref{stress} that while pressure usually diverges at the horizon, the energy density always remains finite in FP2, and so does the total amount of static energy that can be stored outside the horizon. Therefore, as in the case of superluminal fluids, to have any non-zero rate of accretion the surrounding screening matter should freely fall at horizon and hence the pressure is expected to remain finite. This seems to be the reason for the non-uniqueness of the disappearance rate. The rate will then depend on the motion of the collapsing star according to one of the following scenarios:

i) If the total amount of negative energy inside static stars diverges with their redshift (similar to the divergence of total screening charge in the interior of the charged shell in massive electrodynamics), then by slowly contracting the star to higher redshifts we obtain arbitrary rates of accretion. This implies that the gravitational radius $r_g$ (which is approximately the actual radius of the star) shrinks as the redshift increases, and the perceived mass of the star decreases. In this case if the process of collapse is paused at some point the result will be a static star of smaller gravitational radius.

ii) Otherwise, as in the case of impenetrable stars surrounded by superluminal fluids, non-zero accretion rates are obtained when the surface freely falls which allows for the accumulation of an arbitrary amount of free-falling screening matter. As before, the gravitational radius (and the real radius) of the system shrinks in this process. However, if we decide to pause the collapse, the excess of screening matter must return to larger radii, resulting in an excreting star with ever increasing surface redshift.

We were unable to find high redshift static star solutions in FP2 to decide between the two scenarios. Nevertheless, in both cases as the star shrinks to very small radii the quantum mechanical effects become important and the above description in terms of stars breaks down. The resulting small mass black hole will probably evaporate via Hawking radiation. 

%%%%%%%%%%%%%%%%%%%%%%%%%%%%%%%%%%%%%%%%%%%%%%%%%%%%%%%%%%%%%%
\subsection{\label{ADM}A substitute for the ADM mass}

Let us conclude by a discussion of global energy in asymptotically flat solutions of massive gravity. We have already seen that the invariant ADM mass of a localized system trivially vanishes in massive gravity because of the Yukawa screening. Equivalently, if one defines an ordinarily conserved pseudo-tensor of stress and energy, say $\mathfrak{T}^\nu_\mu$, the conserved total energy-momentum four-vector of the system which is obtained by integrating $\mathfrak{T}^0_\mu$ over the whole space is always zero. Intuitively, the negative contribution of $S_{FP}$ cancels the energy and momentum of the `matter content', by which we mean everything except $S_{FP}$. Of course, on a curved space-time these different contributions are not separately conserved, and seemingly, nothing forbids the growth or disappearance of the matter content at the expense of the growth and disappearance of the FP content. However, there exist four Noether charges, associated with four global symmetries of the Fierz-Pauli theory, which as we will see are closely related to the negative energy-momentum of $S_{FP}$. As such, they are also a well-defined measure of the matter-content of the theory. The conservation of these charges, then, constrain processes that involve stars, but they can be violated in the presence of black holes.\footnote{We thank Gregory Gabadadze for discussions and collaboration that lead to the identification of these charges.}

The prescription \eqref{H} to make $S_{FP}$ covariant, also introduces four global symmetries under constant shifts of the scalar fields $\phi^a$. Therefore, there are four conserved Noether currents obtained by varying $S_{FP}$ with respect to $e^a_\mu=\d_\mu\phi^a$
\bea
J^\mu_a= \frac{1}{\sqrt{-g}}\frac{\delta S_m}{\delta e^a_\mu},
\eea
and associated to them, there will be four conserved charges $Q_a$. Moreover, the form of the Lagrangian $U(g^{\mu\nu}e^a_\mu e^b_\nu \eta_{cb})$ allows us to relate $J^\mu_a$ to the stress-tensor of $S_{FP}$
\bea
\label{JT}
J^\mu_a = e^\nu_a(T^\mu_\nu+m^2\delta^\mu_\nu U),
\eea
where $T_{\mu\nu}=2(-g)^{-1/2}\delta S_{FP}/\delta g^{\mu\nu}$, and $e^\mu_a$ is the inverse of $e^a_\mu$. In the unitary gauge $e^a_b=\delta^a_b$, and therefore, $e^a_\mu$ is invertible as long as this gauge exists. Consider now a localized material system whose center of mass is at rest; $Q_0$ has the following properties:

i) When the matter distribution is so dilute that the perturbations of the unitary-gauge metric are infinitesimal ($h_{ab}\ll 1$), we have $J^a_b=T^a_b+\O(h_{ab}^2)$. It follows that $Q_0=-M$, where $M=\int d^3\mbf{r}\T^0_0$ is the total energy of the matter distribution. This can be seen from the fact that when linearization is possible, $\mathfrak{T}^\nu_\mu$ is simply the total stress-energy tensor appearing on the right hand side of the Einstein equation, so we have
\be
M_{\rm ADM}=\int d^3\mbf{r}\mathfrak{T}^0_0=\int d^3\mbf{r}(T^0_0+\T^0_0)=0.
\ee
The condition $h_{ab}\ll 1$ is satisfied as long as $G\T^0_0 \ll m^2$ and $GM^{2/3}{\T^0_0}^{1/3}\ll 1$.

ii) Inside the Vainshtein region, the gravitational field of a star is the same as in the Einstein theory. For a star at finite redshift one can show that $M=-Q_0$, where $M=4\pi r_g$ is the Schwarzschild mass of the star as determined from its Einsteinian field: 

\vspace{-0.3cm}

\begin{quote}
{\it Consider a sphere of radius $r_1$ where $r_g\ll r_1\ll r_V=(r_g/m^2)^{1/3}$. The space-time is already nearly flat at $r_1$, so the ADM mass inside the sphere is well-defined and equal to $M$. The ADM mass inside a much larger sphere of radius $r\to \infty$ is zero, so we must have 
\be
\label{intT0}
\int_{r_1}^\infty d^3\mbf{r}T^0_0=-M.
\ee
In appendix \ref{Q0} we show that 
\be
\label{intJ0=T0}
\int_{r_1\gg r_g}^\infty d^3\mbf{r}J^0_0=\int_{r_1\gg r_g}^\infty d^3\mbf{r}T^0_0.
\ee
Moreover, the assumption of finite redshift ensures that $J^0_0=\O(m^2)$ and the contribution of the region $r<r_1$ to $Q_0$ is negligible, because $J^0_0 r_1^3\ll J^0_0 r_V^3=\O(r_g)$. This fact together with equations \eqref{intT0} and \eqref{intJ0=T0}, implies $M=-Q_0$.
}
\end{quote}

iii) If the redshift of the star diverges as inverse powers of $m r_g$, large amounts of screening matter (presumably with negative energy) can be accumulated in the region $r\sim r_g$, in which case the Schwarzschild mass of the star decreases and deviates from $Q_0$.  

iv) When a black hole forms and evaporates conservation of global charges such as $Q_0$ can be violated. Therefore, the vanishing of the asymptotically defined $M_{\rm ADM}$ in massive gravity allows the post black hole matter content to be less than the initial one.

\section*{Acknowledgments}

We are grateful to Lasha Berezhiani, Sergei Dubovsky, Gia Dvali, Gregory Gabadadze, and Matthew Kleban for many useful discussions. We also thank Stanley Deser and Michael Volkov for useful comments. The work of MM was supported by the NASA grant NNX12AF86G S06.

%%%%%%%%%%%%%%%%%%%%%%%%%%%%%%%%%%%%%%%%%%%%%%%%%%%%%%%%%%%%%%%%%%%%
\appendix

\numberwithin{equation}{section}

\section{Discharge mode in massive electrodynamics}

We numerically solve the eigenvalue problem
\be \label{mode1}
(-({d\over d\rho})^2+\gamma ^2+V)\Psi=0, ~~~{d\Psi \over d\rho}=-\gamma \Psi |_{-\infty}, ~~~\Psi =0|_{+\infty}.
\ee
We take $\Psi (\rho_1)=e^{-\gamma \rho_1}$, ${d\Psi \over d\rho}(\rho_1)=-\gamma e^{-\gamma \rho_1}$, where $\rho_1$ is an arbitrary (but negative and large in absolute value) number. We then integrate eq.(\ref{mode1}) from $\rho_1$ to large positive $\rho$. We choose $\gamma$ so as to get $\Psi \rightarrow 0 ,~~\rho\rightarrow +\infty$. This gives the decay rates $\gamma$. 

The resulting decay rate, for $mr_g<0.5$, to better than 3\% accuracy, is given by a fitting formula 
\be
\gamma \approx {m^2r_g\over 1+mr_g}.
\ee

The asymptotic
\be
\gamma \rightarrow m^2r_g, ~~~mr_g\rightarrow 0
\ee
can be shown to be exact. One first calculates the (singular at horizon) stationary electric field from the Proca equation (\ref{proca3})
\be
{d\over dr}\left ((1-{r_g\over r}){1\over r^2}{d\over dr}(r^2E)\right)=m^2E.
\ee
It follows that
\be\label{exact1}
{d\Psi \over d\rho}+(1-{r_g\over r}){1\over r}\Psi=-m^2r\int _{r}^\infty dr E.
\ee
For small $m$, the right-hand side of eq.(\ref{exact1}) can be approximated by the zero-mass solution $E=q/r^2$, giving
\be\label{exact12}
{d\Psi \over d\rho}+(1-{r_g\over r}){1\over r}\Psi=-m^2q.
\ee
For $r$ close to the horizon, $r-r_g\ll r_g$, we get
\be\label{exact2}
{d\Psi \over d\rho}=-m^2q.
\ee
The decay mode, for $r-r_g\ll r_g$, is $\Psi =Ce^{-\gamma \rho}$ where $C$ is some constant, so that 
\be\label{exact3}
{d\Psi \over d\rho}=-\gamma \Psi.
\ee
On the other hand, for negative $\rho$ in the interval $1\ll |\rho|\ll \gamma ^{-1} $, the decay mode and the zero-mass static mode are approximately equal to the Coulomb field at horizon, $\Psi \approx q/r_g$, and comparing eqs.(\ref{exact2}, \ref{exact3}) we get $\gamma =m^2r_g$.

%%%%%%%%%%%%%%%%%%%%%%%%%%%%%%%%%%%%%%%%%%%%%%%%%%%%%%%%%%
\section{\label{static}Static spherically symmetric field}

In this appendix we outline the method used to find the static solution and a few relevant results of \cite{stars}. Starting from the unitary gauge, the asymptotically flat metric can be written as 
\be 
ds^2=e^\nu dt^2-e^{\tilde{\lambda }}dR^2-R^2e^\mu d\Omega ^2.
\ee
We then change the radial coordinate $R$ to the circumference defined $r$ 
\be 
\label{sch_gauge}
ds^2=e^\nu dt^2-e^\lambda dr^2-r^2d\Omega ^2.
\ee
The scalar fields in the new coordinates become
\be
\label{phi}
\phi^0=t\,,\qquad \phi^i=re^{-\mu/2}n^i\,,
\ee
where $n^i$ is the unit radial vector. 

As independent vacuum equations we use two Einstein equations and the stress-energy conservation
\be \label{e0}
G^0_0=r^{-2}(1-e^{-\lambda})+r^{-1}e^{-\lambda}\lambda '=T^0_0, 
\ee
\be \label{e1}
G^1_1=r^{-2}(1-e^{-\lambda})-r^{-1}e^{-\lambda}\nu '=T^1_1, 
\ee
\be \label{eb}
{T^1_1}'=\frac{1}{2}\nu'(T^0_0-T^1_1)+\frac{2}{r}(T^2_2-T^1_1),
\ee
where prime denotes the $r$-derivative. Inside stars one adds the matter stress-energy tensor $\T^\nu_\mu$ to the right hand side of eqs.(\ref{e0},\ref{e1}), and separately imposes its conservation $\nabla_\nu \T^\nu_\mu =0$.

The stress-energy tensor of FP2 is derived from eq.\eqref{FP2}: 
\be \label{t0}
T^0_0=-m^2(\lambda _1+2\lambda_2+c_2(2\lambda _1\lambda _2+\lambda_2^2)+c_3\lambda_1\lambda_2^2),
\ee
\be \label{t1}
T^1_1=-m^2(\lambda _0+2\lambda_2+c_2(2\lambda _0\lambda _2+\lambda_2^2)+c_3\lambda_0\lambda_2^2),
\ee
\bea \label{t2}
 T^2_2=-m^2(\lambda _0+\lambda_1+\lambda_2  +
 c_2(\lambda _0\lambda _1+\lambda _0\lambda _2+\lambda _1\lambda _2)+c_3\lambda_0\lambda_1\lambda_2), \nonumber
\eea
with 
\be
c_2=1+\tilde c_2,~ c_3=\tilde c_2 +\tilde c_3,
\ee
and
\be 
\label{l0}
\lambda _0=1-e^{-\nu/2},
\ee
\be
\label{l1}
\lambda _1=1-e^{-\tilde{\lambda }/2}=1-e^{-\lambda/2}(1-\lambda_2-r\lambda_2'),
\ee
\be
\label{l2}
\lambda _2=1-e^{-\mu/2}.
\ee
Using expressions (\ref{t0},\ref{t1},\ref{t2}) and expression (\ref{l0},\ref{l1}) in eqs.(\ref{e0},\ref{e1},\ref{eb}) gives the following system of three equations for three unknowns $\nu$, $\lambda$, and $\lambda _2$:
\bea \label{e0n}
1-e^{-\lambda}+re^{-\lambda}\lambda '=-m^2r^2( 2\lambda _2 
+ c_2\lambda_2^2+ 
(1+2c_2\lambda _2+c_3\lambda_2^2)(1-e^{-\lambda /2}(1-\lambda_2-r\lambda_2'))) ~~~
\eea
\bea \label{e1n}
1-e^{-\lambda}-re^{-\lambda}\nu '=-m^2r^2( 2\lambda _2+c_2\lambda_2^2+
(1+2c_2\lambda _2+c_3\lambda_2^2)(1-e^{-\nu /2}) )~~~
\eea
\bea \label{ebn}
r\nu '(1+2c_2\lambda _2+c_3\lambda_2^2)=4(e^{\lambda /2}-1)( 1+c_2\lambda_2+
 (c_2+c_3\lambda _2)(1-e^{-\nu /2}) ).~~~
\eea

After linearizing in $\nu$, $\lambda$, $\mu$, equations (\ref{e0n}, \ref{e1n}, \ref{ebn}) can be solved exactly:
\bea 
\label{linear}
\nu =-ce^{-mr}{1\over r}, ~~\lambda ={c\over 2}e^{-mr}({1\over r}+m),\\ \nonumber 
\mu = {c\over 2m^2}e^{-mr}({1\over r^3}+{m\over r^2}+{m^2\over r})\,,
\eea
manifesting the vDVZ discontinuity. When $mr_g\ll 1$, we have $c=4r_g/3$ where $r_g=M/4\pi$.

Without linearizing, equations (\ref{e0n}, \ref{e1n}, \ref{ebn}) cannot be solved as written. Although we do have three equations for the three unknowns, it is seen that the only derivatives of the unknowns in these equations are $\lambda '$ and $\lambda _2 '$ in eq.(\ref{e0n}), $\nu '$ in eq.(\ref{e1n}), and $\nu '$  in eq.(\ref{ebn}). What makes it possible to solve the system is to derive an algebraic relation between $r$, $\nu$, $\lambda$, $\lambda _2$ by equating the two expressions for $\nu '$. Then one can select any two of the three unknowns $\nu$, $\lambda$, $\lambda _2$ and derive a system of two first-order differential equations for the two selected unknowns.

However, even in the current form one can use Vainshtein approximation to extract a solution which is valid well inside the Vainshtein radius $r\ll r_V = (r_g/m^2)^{1/3}$, but before getting too close to the horizon, where $\l_0$ diverges as $(1-r_g/r)^{-1/2}$ and $T^\mu_\nu \sim m^2\l_0$ can no longer be neglected from the Einstein equations. 

When $r_g\ll r\ll r_V$, we substitute the linearized Schwarzschild solution $\nu=-\lambda=-r_g/r$ into \eqref{ebn}, also linearized in $\nu$ and $\lambda$, but exact in  $\lambda_2$, and solve for $\lambda_2$. We find
\bea
\label{1/c3}
\lambda_2^2=1/c_3\,.
\eea
which implies $c_3>0$. Numerical integration shows that the positive root is connected to the asymptotically decaying solution. 

When $r\to r_g$ but still $G m^2 (1-r_g/r)^{-1/2}\ll 1$, and assuming that $\l_1$ and $\l_2$ remain finite we expect the geometry to be close to Schwarzschild. Substituting $e^\nu=e^{-\lambda}=1-r_g/r$ in \eqref{ebn} and solving for $\lambda_2(r)$ gives, at $r=r_g$\footnote{The larger root was chosen for $\lambda_2$ because in the ``Vainshtein region'' $\lambda_2=1/\sqrt{c_3}$ which is larger than both roots in the parameter range of interest $c_2^2\leq c_3$.}:
\bea
\label{lambda2}
&\lambda_2=-2-c_2/c_3 +\sqrt{4+c_2^2/c_3^2-1/c_3}\,,&\\
&r_g (1-r_g/r)^{1/2}\lambda_2'=-\lambda_2-2c_2/\sqrt{4c_3^2+c_2^2-c_3} \,.
\label{lambda2'}
\eea
The same substitution $e^{-\lambda}=1-r_g/r$ in \eqref{l1} gives, at $r=r_g$,
\bea
\label{lambda1}
\lambda_1=1+r_g (1-r_g/r)^{1/2}\lambda_2'\,.
\eea
These a posteriori justify our assumption of $\l_{1,2}$ remaining finite. In the limit $m\to 0$ these approximate solutions are expected to become exact, which is confirmed by numerical integration. 

%%%%%%%%%%%%%%%%%%%%%%%%%%%%%%%%%%%%%%%%%%%%%%%%%%%%%%%%%%
\subsection{\label{stress} Stress-energy components}

The energy density and stresses in various regions of interest can be obtained from the above approximate solutions:

When $r_V\ll r\ll m^{-1}$, it is seen from eqs. \eqref{l1}, \eqref{l2}, and \eqref{linear} that $\l_2\simeq \mu/2$ and $\l_1\simeq \l_2+r\l_2'$ are the largest eigenvalues, so we can approximate
\bea
T^0_0=m^2\mathcal{O}(r_g/r), ~~~~~~~~~\\
T^1_1\simeq -2 T^2_2\simeq -\frac{2}{3}m^2\left(\frac{r_V}{r}\right)^3.
\eea
When $r_g\ll r\ll r_V$, still $\l_1\simeq \l_2\simeq 1/\sqrt{c_3}$ are the larges eigenvalues, and we have
\bea
T^0_0\simeq -m^2(4+3 c_2/\sqrt{c_3})/\sqrt{c_3},~~~~ \\
T^1_1\simeq T^2_2\simeq -m^2(2+ c_2/\sqrt{c_3})/\sqrt{c_3}.
\eea
Finally, when $r\to r_g$ and $\l_0\to -\infty$
\bea
T^0_0=m^2\mathcal{O}(1), ~~~~~~~~~~~~~~~~~~~~~~~~~\\
T^1_1\simeq - m^2 \l_0 (1+2c_2\lambda_2+c_3\lambda_2^2),~~~~~~~\\
T^2_2 \simeq - m^2 \l_0 (1+c_2(\lambda _1+\lambda _2)+c_3\lambda _1\lambda _2),
\eea
where $\l_{1,2}$ are given in \eqref{lambda2} and \eqref{lambda1}. Pressure $p=-(T^1_1+2T^2_2)/3$ diverges at horizon unless $c_2^2=c_3$. Note that since the energy density $T^0_0$ is negative on average, $p\to -\infty$ at horizon should not be very surprising -- reversing the sign of the action does not change the equations of motion, but it reverses the sign of stress-energy tensor. 

%%%%%%%%%%%%%%%%%%%%%%%%%%%%%%%%%%%%%%%%%%%%%%%%%%%%%%%%
\section{\label{app-near}Near-hole solutions in FP2}

As argued in \S\ref{near} the quasi-stationary near-hole solutions can be obtained from $\nabla_\mu T^\mu_\nu=0$, or equivalently, from the equations of motion for $\phi^a$ on Schwarzschild geometry. To derive these equations, we first need to determine the eigenvalues $\lambda_a$ of the matrix $\delta^a_b-\sqrt{H^a_b}$, in terms of which the FP action is defined. For the spherically symmetric metric 
\be
\label{g2}
ds^2=C dT^2-2D dT dR -A dR^2 - B R^2 d\Omega^2.
\ee
they are given by
\bea
\lambda_\pm&=&1-\frac{1}{\sqrt{2\Delta}}\left[A+C\pm\sqrt{(A-C)^2-4D^2}\right]^{1/2},\\
\label{l2C}
\lambda_2&=&\lambda_3=1-B^{-1/2},
\eea
where
\be
\Delta=AC+D^2,
\ee
and we have renamed $\lambda_0=\lambda_+$ and $\lambda_1=\lambda_-$. \footnote{\label{kappa}One can see from these expressions why $T^{01}\propto \delta S_{FP}/\delta D$ (and similarly $T_{01}$) is proportional to $D$ with a regular proportionality coefficient at $D=0$. $S_{FP}$ depends on $D$ via $\lambda_{\pm}$, and $\sqrt{\Delta}$ in the overall $\sqrt{-g}$. Both are quadratic in $D$ except when: (i) $A=C$, in which case since $\lambda_+=\lambda_-$ at $D=0$ the symmetry of $S_{FP}$ in $\lambda_{\pm}$ ensures that terms linear in $D$ cancel. (ii) $AC=0$ which is a singular point at $D=0$ and therefore cannot be approached from $D=0$ side, which is dictated by the asymptotic condition.} Since the FP2 action \eqref{FP2} depends on $\lambda_\pm$ through the combinations $\lambda_++\lambda_-$ and $\lambda_+\lambda_-$, it is useful to define
\bea
F=\frac{1}{\sqrt{2}}\left(\left[A+C +\sqrt{(A-C)^2-4D^2}\right]^{1/2}+\left[A+C -\sqrt{(A-C)^2-4D^2}\right]^{1/2}\right),
\eea
in terms of which  
\bea
\lambda_++\lambda_-&=&2- F/\sqrt{\Delta},\\
\lambda_+\lambda_-&=&1+(1-F)/\sqrt{\Delta}.
\eea
The FP2 Lagrangian becomes
\be
\label{U}
U=-\frac{F}{\sqrt{\Delta}} \beta  +\frac{1}{\sqrt{\Delta}} \tilde \beta+\beta+2\l_2+c_2\l_2^2,
\ee
where we defined the new parameters
\bea
\label{cs}
c_2=1+\tilde c_2,\qquad c_3=\tilde c_2+\tilde c_3,
\eea
and the functions 
\bea
\label{beta}
\beta \equiv 1+2c_2\l_2+c_3\l_2^2,\qquad
\tilde\beta \equiv 1+2\tilde c_2\l_2+\tilde c_3\l_2^2.
\eea
The stress-energy tensor components can now be calculated by varying the action with respect to the metric:
\be
m^{-2}T^\nu_\mu=-2 g_{\mu\sigma}\frac{\d U}{\d g_{\nu\sigma}}-\delta^\nu_\mu U.
\ee
For instance, we get
%%%% using
%%%% \be
%%%% \sqrt{(A-C)^2-4D^2}=\sqrt{\Delta}F(\l_--\l_+).
%%%% \ee
\be
\label{T10}
m^{-2}T^1_0=C\frac{\d U}{\d D}-2D\frac{\d U}{\d A}=\frac{\beta D}{\sqrt{\Delta}F}.
\ee

After deriving the stress-energy components one can use the Vainshtein approximation \eqref{CA}, namely,
\bea
\label{CA1}
C=1-r_g/r,\qquad A=C^{-1}(\gamma^{-2}-D^2),
\eea
to express everything in terms of $r$ and the two unknowns $\l_2$ and $D$. In particular, using $R=B^{-1/2}r$, we get
\be
\frac{1}{\sqrt{\Delta}}=\gamma=dR/dr=1-\lambda_2-r\lambda_2'.
\ee
The equations for $D$ and $\l_2$ can be derived from the stress-energy conservation 
\bea
\label{dT}
\sqrt{-g}\nabla_\mu T^\mu_\nu= \d_\mu(\sqrt{-g}T^\mu_\nu)-\frac{1}{2}\sqrt{-g}T^{\alpha\beta} \d_\nu g_{\alpha\beta}=0,
\eea
which is particularly simple for the $\nu =0$ component in the quasi-stationary approximation. It becomes the condition of steady energy flux, $\partial_1(\sqrt{-g}T^1_0)=0$, whose integration constant is (minus) the accretion rate $\dot r_g=-\alpha m^2r_g^2$. Using the expression \eqref{T10}, we obtain
\bea
\label{D}
\frac{D\beta}{F}=\frac{\alpha r_g^2}{BR^2}=\frac{\alpha r_g^2}{r^2}.
\eea
Another independent equation can be obtained from the $\nu=1$ component of \eqref{dT} which together with the $\nu=0$ component form a linear system of equations for $D'$ and $\l_2''$. However, it is technically easier to derive this second equation by using the Vainshtein approximation directly inside the FP2 Lagrangian \eqref{U} and writing it as an action for the scalar fields $\phi^a$ on Schwarzschild background:
\be
S=\int dr r^2U.
\ee
Here $U$ is a function only of $r$, $D$, $\l_2$, and $\g$. As a consistency check, note that since ${\phi^0}'= D\gamma/C$ the equation of motion for $\phi^0$ is given by
\be
\frac{\delta S}{\delta \phi^0}=-\d_r\left(r^2\frac{C}{\gamma}\frac{\d U}{\d D}\right)=-\d_r\left(\frac{D\beta  r^2}{F}\right)=0,
\ee
which integrates to our first equation \eqref{D}. The equation of motion for $\phi^i=(1-\l_2)r n^i$ can be obtained by varying $S$ with respect to $\l_2$, but remembering that $\phi^0$ also depends on $\l_2$ via $\gamma=1-(r\l_2)'$. Subtracting that contribution, we get
\be
r^2\frac{\d U}{\d \l_2}+r\d_r\left[r^2\left(\frac{\d U}{\d \gamma}-\frac{D}{\gamma}\frac{\d U}{\d D}\right)\right]=0,
\ee
which after some algebra, and using the identity
\be
\label{CF2}
CF^2+D^2 = \g^{-2}(1+\g C)^2,
\ee
yields
\be
\label{l22}
r\d_r\left[\frac{\beta r^2(1+\gamma C)}{\gamma F}-\tilde \beta r^2\right]=
r^2 (1-\gamma F) \frac{\d \beta}{\d \l_2}+r^2 \gamma \frac{\d \tilde \beta}{\d \l_2}+2r^2(1+c_2 \l_2).
\ee
This equation contains both $\l_2''$ and $D'$, but it becomes algebraic in $\l_2$ once we eliminate $D$ using 
\be
\label{golden}
F=(\g^{-1}+ C)\left(C+\frac{\alpha^2r_g^4}{\beta^2 r^4}\right)^{-1/2},
\ee
which can be obtained by substituting $D=\alpha r_g^2 F/r^2\beta$ from \eqref{D} into the identity \eqref{CF2}. After this substitution, the left hand side of \eqref{l22} which would have contained $\l_2''$ and $D'$ becomes
\be
r\d_r\left[(C\beta^2 r^4+\alpha^2 r_g^4)^{1/2}-\tilde \beta r^2\right].
\ee
It is manifestly independent of $\l_2''$. Factors of $\l_2'$ will also cancel from the two sides, and we finally get
\bea
\label{l222}
\beta (1-\frac{3r_g}{4r})+(c_2+c_3\l_2)(1+(1-\l_2)C)=\left(C+\frac{\alpha^2r_g^4}{\beta^2 r^4}\right)^{1/2}(1+c_2\l_2+2(c_2+c_3\l_2)).~~~~~
\eea
For given $\alpha$ and $r_g$, $\l_2$ is solved for from \eqref{l222}, using which $D$ is given by
\be
D=\frac{\alpha r_g^2}{\beta r^2}(\g^{-1}+C)\left(C+\frac{\alpha^2r_g^4}{\beta^2 r^4}\right)^{-1/2}
\ee
which is obtained by substituting \eqref{golden} in \eqref{D}.

%%%%%%%%%%%%%%%%%%%%%%%%%%%%%%%%%%%%%%%%%%%%%%%%%%%%%%%%%%%
\section{\label{trans}Transition region}

It is difficult to derive analytic results for the transition region $r\sim r_V=(r_g/m^2)^{1/3}$, but since accretion effects rapidly disappear away from the horizon one can safely linearize in $D$ and use the knowledge about static solution in this region. After this linearization, the only component of the stress tensor that depends on $D$ is (c.f. eq.\eqref{T10})
\be
\label{T011}
T_0^1\simeq m^2 \frac{D\beta}{\sqrt{\Delta} (\sqrt{A}+\sqrt{C})},
\ee
the rest being given by their $\alpha=0$ expressions. Here $\Delta\simeq AC$ and the Vainshtein approximation \eqref{CA1} is not used, so \eqref{T011} is valid even for $r>r_V$.

Inside the Vainshtein radius $r\ll r_V$, we have, from eq.\eqref{D}, a steady flow with energy flux
\be
\label{flux1}
\sqrt{\Delta}T^1_0=\alpha m^2 \frac{r_g^2}{r^2},
\ee
corresponding to $\dot r_g=-\alpha m^2 r_g^2$, which in turn fixes the linearized solution \eqref{d1} of \S \ref{lin} and from \eqref{T011} we get for $r_V\ll r\ll m^{-1}$
\bea
\label{flux2}
\sqrt{\Delta}T^1_0\simeq m^2\frac{d}{2}=\frac{2}{3}\alpha m^2 \frac{r_g^2}{r^2}.
\eea
So there is a net positive flux of energy into the transition region. This can be understood in the following way. On a static solution, the relevant degree of freedom in this region is $R/r =B^{-1/2}$, or the related quantity $\l_2$ (see eqs.\eqref{l2}, \eqref{linear}, \eqref{1/c3})
\bea
\l_2\simeq\left\{\begin{array}{cc} 
\frac{1}{3}(r_V/r)^3 & r_V\ll r\ll m^{-1},\\
1/\sqrt{c_3} & r_g\ll r\ll r_V,\end{array}\right. 
\eea
which is approximately a function only of the ratio $r/r_V$. The energy density profile is therefore a function of $r/r_V$; as was shown in appendix \ref{stress}, in going from $r\gg r_V$ to $r\ll r_V$ it interpolates between a negligible amount to a negative constant $\vep \sim -m^2$. Once we allow accretion, $r_g$ gradually decreases and as a result the Vainshtein radius and the above-mentioned profile adiabatically shrink to smaller radii. Therefore, the total amount of energy in this region $\Delta M\sim \vep r_V^3 \sim -r_g$ grows, requiring a net positive energy flux. 

The exact value of $\Delta M$ can be found from the difference of the invariant masses $M_1$ and $M_2$ inside two spheres of radii $r_1$ and $r_2$, where $r_g\ll r_1\ll r_V$ and $r_V\ll r_2\ll m^{-1}$. Since at distances $r\gg r_g$ the geometry is nearly flat, the metric can be transformed into $g_{\mu\nu}=\eta_{\mu\nu}+h_{\mu\nu}$ with $h_{\mu\nu}\ll 1$, in terms of which the invariant mass inside a sphere of radius $r$ is given by
\bea
M=\frac{1}{2}\int (\d_i h_{jj}-\d_j h_{ij})n^i r^2 d\Omega.
\eea
Using circumference defined radius to parametrize the static metric [as in \eqref{sch_gauge}], we get $h_{ij}=-\lambda n^i n^j$, where inside the Vainshtein radius $\lambda$ is given by the Einsteinian $r_g/r$ expression, and outside by the vDVZ one $2r_g/3r$. The enclosed masses are then
\be
M_1=4\pi r_g,\qquad M_2 =\frac{8\pi}{3}r_g,
\ee
giving 
\be
\Delta M = M_2-M_1=-\frac{4\pi}{3}r_g.
\ee
Therefore, when $\alpha\neq 0$ the time-dependence of $\Delta M$ exactly accounts for the difference between \eqref{flux1} and \eqref{flux2}. 

Similar methods can be used to find $D$ throughout the transition region. One uses the numerical static solution of \cite{stars} to determine $M(r,r_g)$. By numerically differentiating $M(r,r_g)$ with respect to $r_g$, one can find 
\be
\dot M(r,r_g)=\dot r_g \d_{r_g}M(r,r_g).
\ee
The energy flux at any $r_2\sim r_V$ is determined from
\be
\left. \dot M(r_2)-\dot M_1 = -r^2 \sqrt{\Delta}T^1_0\right|_{r_1}^{r_2},
\ee
using which one solves for $D$ from \eqref{T011}.  

%%%%%%%%%%%%%%%%%%%%%%%%%%%%%%%%%%%%%%%%%%%%%%%%%%%%%%%%%%
\section{\label{Q0} The mass of a star and the global charge $Q_0$}

Consider a star of gravitational radius $r_g$ deep inside the Vainshtein regime. In the Schwarzschild coordinates, where the metric is given by
\be
\label{g1}
ds^2 = e^\nu dt^2-e^\lambda dr^2-r^2d\Omega^2,
\ee
we have $e^\nu=e^{-\l}=1-r_g/r$ for $r\ll r_V$. We will show in what follows that for finite redshift stars the Noether charge $Q_0$ coincides with minus the Schwarzschild mass $4\pi r_g$ of the star.

Note first that for all $r\gg r_g$ the functions $\nu$ and $\l$ are infinitesimal (although different from the Einsteinian value when $r\gg r_V$), so that the space-time is nearly flat. Thus, the ADM mass within any sphere of radius $r\gg r_g$ is well-defined. Consider a sphere of radius $r_1$ where $r_g\ll r_1\ll r_V$. The enclosed ADM mass coincides with the Schwarzschild mass $4\pi r_g$. On the other hand the ADM mass inside a sphere of much larger radius, $r\gg m^{-1}$, must vanish in massive gravity; therefore, the contribution of $S_{FP}$ to $M_{\rm ADM}$ coming from the region $r_1<r<\infty$ should be exactly $-4\pi r_g$. Given the closeness of the metric \eqref{g1} to Minkowski in this region, this contribution of $S_{FP}$ can be directly calculated from the volume integral of its energy density, namely
\be
\int^{\infty}_{r_1\gg r_g }d^3\mbf r T^t_t=-4\pi r_g.
\ee
We will show that the integral on the left coincides with the volume integral of $J^t_0$ in the same region of space, which is what one expects if $Q_0$ is a good characterization of the energy of $S_{FP}$:

Since for $r\gg r_g$ the unitary-gauge metric is approximately diagonal (even if there is time-dependence), we have $\phi^0\simeq t$ and $\d_t \phi^i\simeq 0$, so $J^t_0$ and $T^t_t$ can be replaced with the unitary-gauge expressions $J^0_0$ and $T^0_0$. From \eqref{JT} we have (after redefining $m^2U\to U$)
\be
J^0_0= T^0_0-U,
\ee
and the above requirement on $Q_0$ implies
\be
\label{intU}
\int^{\infty}_{r_1\gg r_g}dr r^2 U \ll r_g.
\ee
This is trivially satisfied in the linear regime $r\gg r_V$, where the perturbations of the unitary-gauge metric $h_{ab}$ are themselves infinitesimal, and $U=\O(m^2 h_{ab}^2)$. 

Now consider the contribution from a region $r_1<r<r_2$ where $r_g\ll r_1\ll r_V\ll r_2\ll m^{-1}$. The static solution is a good approximation throughout this region and we can use the formalism of appendix \ref{static} to verify eq.\eqref{intU}. Since $\nu \ll 1$, the stress-energy conservation \eqref{eb} simplifies to
\be
\label{approxeb}
{T^1_1}'\simeq\frac{2}{r}(T^2_2-T^1_1),
\ee
and for a diagonal unitary-gauge metric, $T^1_1$ and $T^2_2$ of a generic FP (including FP2) are given by
\be
\label{T1T2}
T^1_1=(\l_1-1)\frac{\d U}{\d \l_1}-U,\qquad T^2_2=\frac{1}{2}(\l_2-1)\frac{\d U}{\d \l_2}-U,
\ee
(the factor of $1/2$ in $T^2_2$ is because $\l_2=\l_3$). Moreover, in this region $\l\ll \l_2$, and the expression \eqref{l1} for $\l_1$ can be approximated as
\be
\label{approxl1}
\l_1\simeq (\l_2r)',
\ee
while $\l_0\simeq \nu/2$ and its derivatives can be neglected. Substituting \eqref{T1T2} and \eqref{approxl1} in \eqref{approxeb} yields
\be
\label{dUdl2}
r\frac{\d U}{\d\l_2}=\left(r^2\frac{\d U}{\d\l_1}\right)'.
\ee
We can now transform
\bea
\left.\int_{r_1}^{r_2}dr r^2 U=\frac{1}{3}r^3U\right|_{r_1}^{r_2}
-\frac{1}{3}\int_{r_1}^{r_2}dr r^3U'~~~~~~~~~~~~~~~~~~~~~~~~~~~~\nonumber\\
\left.=\frac{1}{3}r^3U\right|_{r_1}^{r_2}
-\frac{1}{3}\int_{r_1}^{r_2}dr r^3\left(\frac{\d U}{\d\l_2}\l_2'+\frac{\d U}{\d\l_1}(\l_2r)''\right),
\eea
and use \eqref{dUdl2} to obtain
\be
\left.\int_{r_1}^{r_2}dr r^2 U=\frac{1}{3}r^3\left(U-r\l_2'\frac{\d U}{\d\l_1}\right)\right|_{r_1}^{r_2}.
\ee
But $r_2$ is in the linear regime where $\l_a\sim h_{aa}\ll 1$, and hence the right hand side which is quadratic in $\l_a$ is negligible at the upper limit. At the lower limit, since $U\propto m^2$, $\l_{1,2}\sim 1$, and $r_1^3\ll r_V^3=r_g/m^2$, we get the desired relation \eqref{intU}.

For a star of finite redshift all eigenvalues $\l_a$ remain finite and the contribution of the region $r < r_1$ to $Q_0$, which is of order $r_1^3J_0^0\sim m^2 r_1^3$, is much less than $r_g$. Therefore, $Q_0$ is given by the total amount of negative mass in the region $r\gg r_g$ evaluated by the volume integral of $T^0_0$ and we have $Q_0=-4\pi r_g$. 

%%%%%%%%%%%%%%%%%%%%%%%%%%%%%%%%%%%%%%%%%%%%%%%%%%%%%%%%%%
\section{\label{fluid} Fluid accretion}

Potential flow of ideal fluids can be described by a single scalar field $\phi$ with the k-essence action \cite{Mukhanov}
\be
S = \int d^4x \sqrt{-g} P(X),
\ee
where $X= g^{\mu\nu} \d_\mu \phi \d_\nu \phi$. The stress-energy tensor is
\be 
\T_{\mu\nu} = 2 P_X \d_\mu\phi \d_\nu \phi -g_{\mu\nu} P,
\ee
where $P_X = \d P/\d X$. We can therefore identify the fluid pressure $p$, energy density $\vep$, and velocity $u_\mu$ as
\be
p = P, \qquad \vep= 2 P_X X -P, \qquad u_\mu =\d_\mu \phi /\sqrt{X}.
\ee
The $p=c_s^2 \vep$ equation of state, therefore, corresponds to
\be
P=X^n,\qquad\text{with}\qquad n=\frac{1+c_s^{-2}}{2}.
\ee

%%%%%%%%%%%%%%%%%%%%%%%%%%%%%%%%%%%%%%%%%%%%%%%%%%%%%%%%%%%
\begin{figure}[t]
\begin{center}
\includegraphics[width=8.cm]{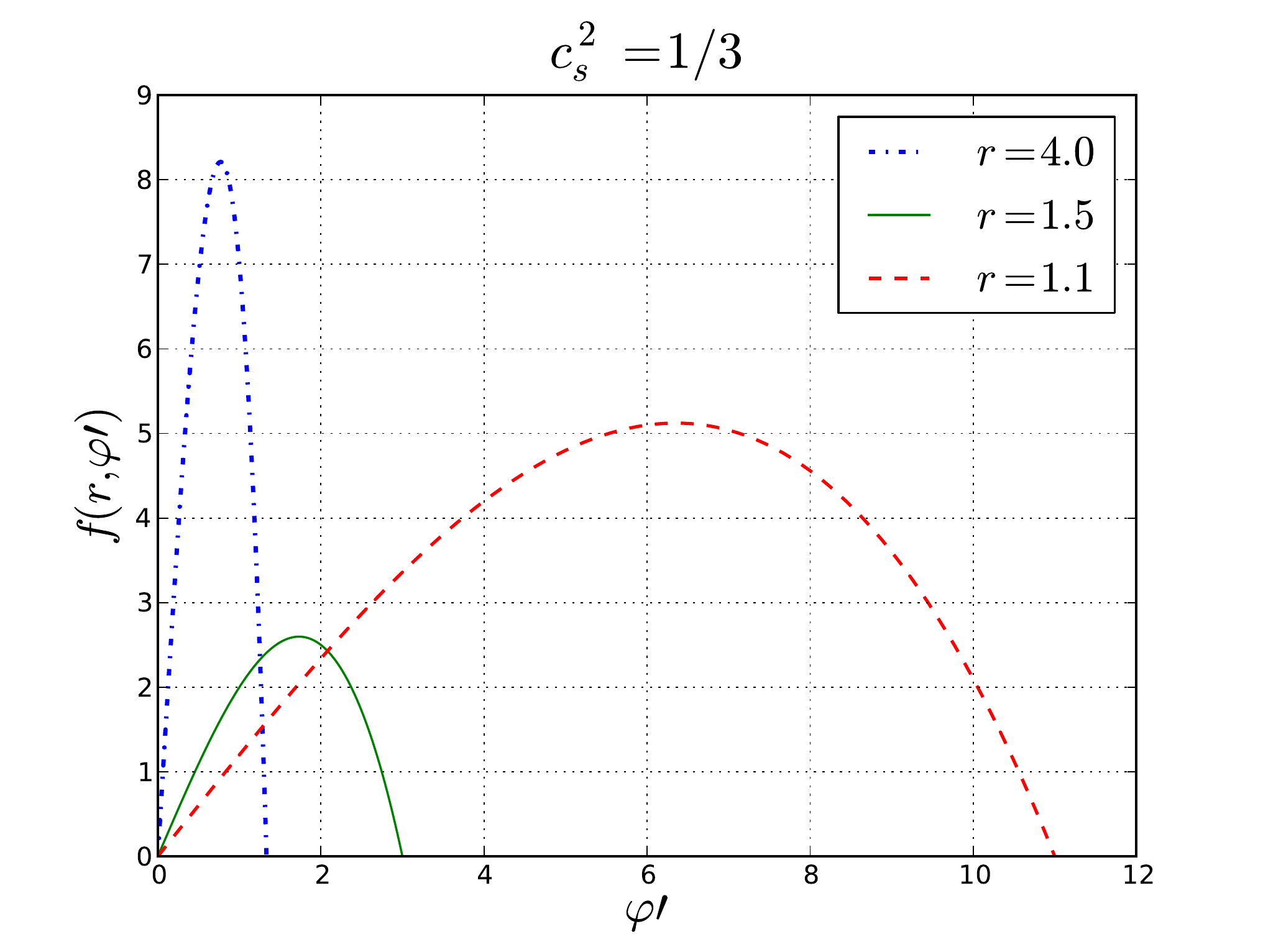}
\includegraphics[width=8.cm]{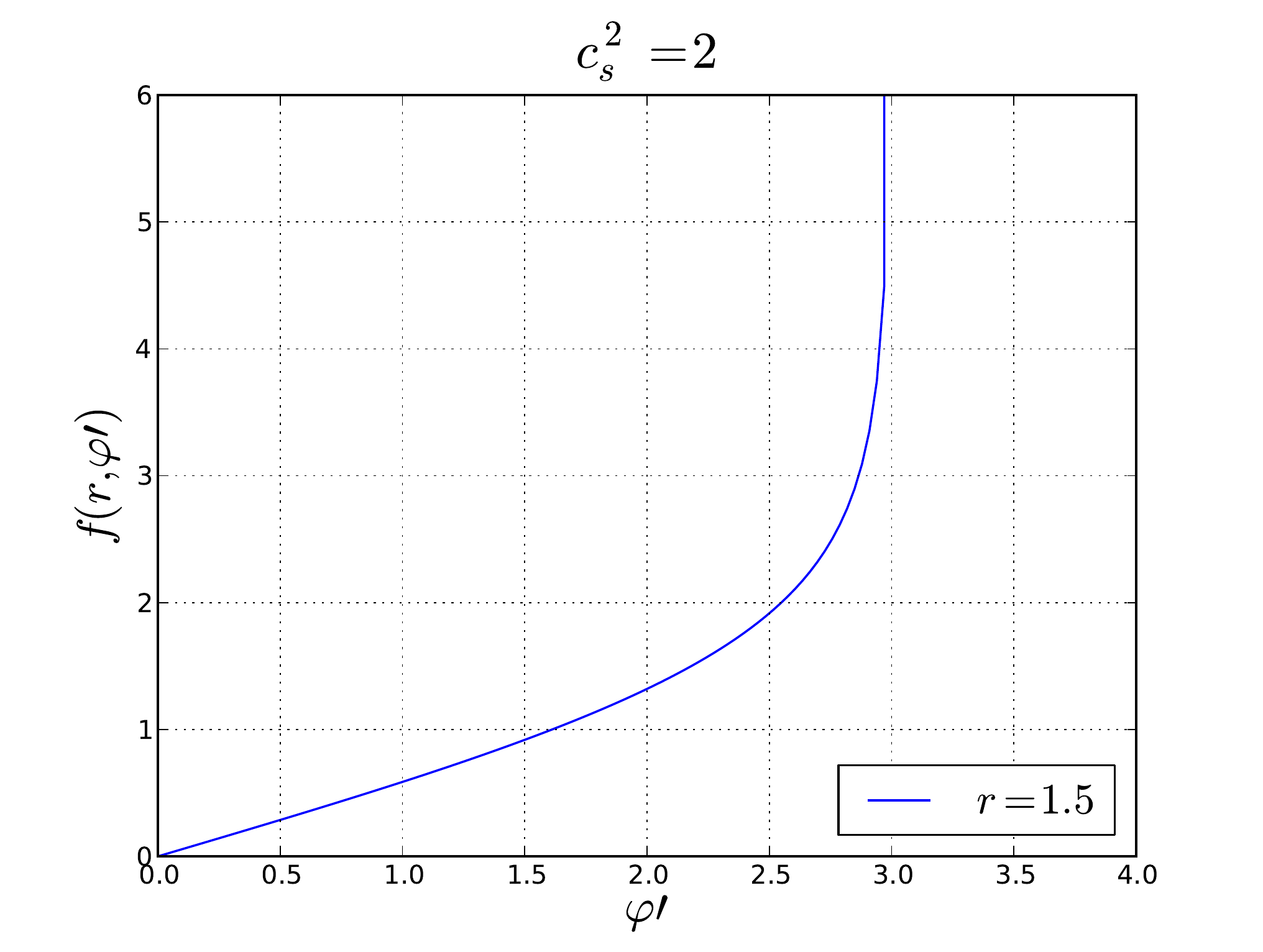}
\caption[]{\small{$f(r,\varphi')$ plotted as a function of $\varphi'$ at fixed values of $r$ and for $r_g=1$. When $c_s^2< 1$ (left), the maximum of $f(r,\varphi')$ decreases until a critical radius is reached and then increases. When $c_s^2>1$ (right), $f(r,\vphi')$ has no maximum.} }

\label{fplot}

\end{center}
\end{figure}
%%%%%%%%%%%%%%%%%%%%%%%%%%%%%%%%%%%%%%%%%%%%%%%%%%%%%%%%%%%%%%%%

For static spherically symmetric fluids with asymptotic density $\vep_0$, we have $\phi = \phi_0 t$, with $\phi_0=(c_s^2 \vep_0)^{1/(1+c_s^{-2})}$, which on Schwarzschild geometry gives the singular at horizon solution \eqref{density}. To have an accreting solution we take $\phi = \phi_0 (t+\vphi(r))$. Integrating once the $\phi$ equation of motion, we get
\be
\label{f}
f(r,\vphi')\equiv r^2(1-\frac{r_g}{r})\vphi'[(1-\frac{r_g}{r})^{-1}-(1-\frac{r_g}{r})\vphi'^2]^{n-1}=A,
\ee
where $A$ is an integration constant. For any $A$, there exists at large values of $r$ a decaying solution $\vphi' \approx A/r^2$, corresponding to the steady accretion
\be
\T_t^r=-(1+c_s^2)\vep_0\frac{A}{r^2}.
\ee
However, when $c_s^2\leq 1$ only one of the decaying solutions matches a regular solution at horizon (the Bondi solution). This can be shown as follows. At any fixed $r$ the function $f(r,\vphi')$ in \eqref{f} is maximized at 
\be
\vphi'=c_s (1-\frac{r_g}{r})^{-1},
\ee
giving a maximum $f_{\rm{max}}(r)$. As $r$ is decreased from infinity, $f_{\rm{max}}(r)$ reaches a minimum at the critical radius
\be
r_c= \frac{3+c_s^{-2}}{4} r_g,
\ee
and then increases (figure \ref{fplot}). For $A> A_{\rm{Bondi}}=f_{\rm{max}}(r_c)$, the asymptotically decaying solution ceases to exist below some radius $r(> r_c)$. For $A< A_{\rm{Bondi}}$, there exist a solution all the way to the horizon but $\vphi'$ is not large enough there. Therefore, $X$ diverges as $(1-r_g/r)^{-1}$, leading to a pressure singularity. Only for $A=A_{\rm{Bondi}}$ the asymptotically decaying solution connects to a rapidly falling solution at horizon, where $\vphi'\to (1-r_g/r)^{-1}$ and $X$ remains finite. 

When $c_s^2>1$, $f(r,\vphi')$ is unbounded above and there is no critical point as can be seen from figure \ref{fplot}. For all $A$ the asymptotically decaying solution goes into
\be
\vphi'=(1-\frac{r_g}{r})^{-1}-\frac{1}{2}\left(\frac{A}{r_g^2}\right)^{-2/(1-c_s^{-2})},
\ee
in the $r\to r_g$ limit, and has a finite pressure. The energy of fluid elements at horizon reaches the constant
\be
e=u_0=\left(\frac{A}{r_g^2}\right)^{1/(1-c_s^{-2})}.
\ee

%%%%%%%%%%%%%%%%%%%%%%%%%%%%%%%%%%%%%%%%%%%%%%%%%%%%%%%%%%%%%%%%%%

\end{document}